\documentclass[%
preprintnumbers,
nofootinbib,
 amsmath,amssymb,
 aps,
]{revtex4-2}

\usepackage{graphicx,color}% Include figure files

\usepackage{dcolumn}% Align table columns on decimal point
\usepackage{bm}
\usepackage{mathtools}
\usepackage{here}

\begin{document}

% Use the \preprint command to place your local institutional report
% number in the upper righthand corner of the title page in preprint mode.
% Multiple \preprint commands are allowed.
% Use the 'preprintnumbers' class option to override journal defaults
% to display numbers if necessary
\preprint{KEK-TH-2574 and J-PARK-TH-0296}

%Title of paper
\title{Is gauge symmetry vacuous or physical ?  :   
Lessons from the Landau problem as a solvable quantum gauge theoretical system}

% repeat the \author .. \affiliation  etc. as needed
% \email, \thanks, \homepage, \altaffiliation all apply to the current
% author. Explanatory text should go in the []'s, actual e-mail
% address or url should go in the {}'s for \email and \homepage.
% Please use the appropriate macro foreach each type of information

% \affiliation command applies to all authors since the last
% \affiliation command. The \affiliation command should follow the
% other information
% \affiliation can be followed by \email, \homepage, \thanks as well.

\author{Masashi Wakamatsu}
\email[]{wakamatu@post.kek.jp}
%\homepage[]{Your web page}
%\thanks{}
%\altaffiliation{}
\affiliation{KEK Theory Center, Institute of Particle and Nuclear Studies, High Energy Accelerator
Research Organization (KEK), 1-1, Oho, Tsukuba, Ibaraki 305-0801, Japan}

%Collaboration name if desired (requires use of superscriptaddress
%option in \documentclass). \noaffiliation is required (may also be
%used with the \author command).
%\collaboration can be followed by \email, \homepage, \thanks as well.
%\collaboration{}
%\noaffiliation

%\date{\today}

\begin{abstract}
The gauge symmetry is one of the most important concepts in modern physics,
but there are two conflicting views on its meaning or interpretation.
The standard view is that local gauge symmetry is the basis of the pursue of 
fundamental particles and forces in nature.
Another view is that the gauge symmetry is not a symmetry of nature but
just a redundancy in description. Naturally, both statements are nothing wrong, 
but one might feel that there is a slight conceptual conflict 
between the two points of view. Due to the subtlety of the subject, however,
little literature exists that discusses the root of such an anxiety.
In the present paper, by making full use of the analytically solvable nature
of the quantum Landau problem, we argue that the familiar gauge principle
plays a critical role in unraveling a subtle mismatch between the two viewpoints above.
We reveal that there exist two types of quantities in gauge theories, which 
should clearly be discriminated.
The first are quantities, which look seemingly gauge-invariant but actually not,
whereas the second are genuinely gauge-invariant quantities, which correspond 
to direct experimental observables.
\end{abstract}

% insert suggested PACS numbers in braces on next line
% 03.65.-w : Quantum mechanics
% 11.15.-q : Gauge field theories
% 11.30.-j : Symmetry and conservation laws
% 12.20.-m : Quantum electrodynamics
%% 12.38.-t : Quantum Chromodynamics
% % 14.20.Dh : Protons and neutrons
%% 52.40.Db : Electromagnetic
% 71.70.Di : Landau levels
% 87.50.Mn : Magnetic fields

%\maketitle must follow title, authors, abstract, \pacs, and \keywords
\maketitle

% body of paper here - Use proper section commands
% References should be done using the \cite, \ref, and \label commands
%\section{}
% Put \label in argument of \section for cross-referencing
%\section{\label{}}
%\subsection{}
%\subsubsection{}
%

%%%%%%%%%%%%%%%%%%%%%%%%%%%
%%%%%%%%%%%%%%%%%%%%%%%%%%%
%%%%%%%%%%%%%%%%%%%%%%%%%%%

\section{Introduction}
\label{sec1}

Gauge symmetry is one of the most important concepts in modern physics,
but there are two conflicting views on its interpretation or meaning.
The widely-accepted standard view would be the following.
In \cite{Gross1997}, Gross says ``We now suspect that all fundamental 
symmtries are local gauge symmetries.'' 
Similarly, Wilczek says in \cite{Wilczek2013}
``A central pillar of the standard model is the idea of gauge symmetry."
However, we point out that there also exist somewhat critical view on the
gauge symmetry concept. For example, in \cite{Schwartz2014}, 
Schwartz says ``Gauge invariance is
not physical. It is not observable and not a symmetry of nature."
Similarly, Zee says in \cite{Zee2013} ``Gauge symmetry is strictly 
speaking not a symmetry but a redundancy in description."
Naturally, both viewpoints are nothing wrong, but it is also true that
a slight conflict between the two viewpoints above raises a nontrivial deep 
question about the role and the meaning of the gauge symmetry concept
in nature.
This is an extremely delicate question, which would make us easily trapped
in a fuzzy or philosophical dispute. To avoid such a trouble, we believe that
discussions on this issue should be done by using some concrete examples,
which are easy for everyone to understand. 
As we shall demonstrate in the present paper, 
there is a particularly useful physical problem for this purpose.
It is the analytically solvable Landau problem as a quantum gauge theoretical
system.

As is widely known, in the traditional formulation of the Landau problem, 
there are three typical choices of gauge, i.e. the symmetric gauge, and 
the 1st and 2nd Landau gauges \cite{Landau1930, Landau-Lifschitz,
Vallentine1998, VGW2006, Goerbig2009, Tong2016}.  
In the symmetric gauge, the Landau eigen-states
are characterized by two quantum numbers $n$ and $m$, where $n$ is
the so-called Landau quantum number characterizing the eigen-energies
of the Landau Hamiltonian, while $m$ is the eigen-value of the
canonical orbital angular momentum (OAM) operator $\hat{L}^{can}_z$. 
In a series of papers \cite{WKZ2018, KWZZ2020, WKZZ2020,WKZZ2021}, 
we tried to demonstrate that the quantum number $m$ does not 
correspond to direct observables in conformity with the fact that the canonical 
OAM is standardly believed to be a {\it gauge-variant} quantity.
Those were arguments after the choice of the symmetric gauge, but we 
naturally anticipate that the similar holds true also in 
other gauges, i.e. also in the two Landau gauges.
For example, with the choice of the 1st Landau gauge, the eigen-states
of the Landau Hamiltonian are characterized by two quantum numbers
$n$ and $k_x$, where $n$ is the same quantum number as appearing
in the eigen-states in the symmetric gauge, whereas $k_x$ is the
eigen-value of the canonical momentum operator $\hat{p}^{can}_x$.
Since $k_x$ is the eigen-value of the {\it gauge-variant} canonical
momentum operator $\hat{p}^{can}_x$, we naturally expect that it too 
does not correspond to experimentally measurable quantities. 
In fact, the validity of this reasonable conjecture was extensively
discussed in a recent paper \cite{WH2022}.

However, this plausible understanding reached in \cite{WH2022} was
criticized in a recent paper \cite{Govaerts2023} based on logical development
of  the {\it gauge-potential-independent formulation} of the Landau problem, 
which predicts the existence of three conserved 
quantities \cite{WH2022, Govaerts2023},  i.e. the two conserved momenta
$\hat{p}^{cons}_x$ and $\hat{p}^{cons}_y$ and one conserved
OAM $\hat{L}^{cons}_z$, the conservation of which is guaranteed
independently of the choices of the gauge potential. 
(We point out that these conserved quantities are basically the same
quantities as those called the {\it pseudo momentum} 
and the {\it pseudo OAM} in some recent 
literature \cite{Yoshioka2002, Konstantinou2016, Konstantinou2017,
KWZZ2020}.).
According to the author of the paper \cite{Govaerts2023}
 (see also the earlier literature \cite{Haugset1993} by Haugset et al.), 
the existence of the three types of Landau eigen-states, i.e. the two 
Landau-gauge-type eigen-states and the symmetric-gauge-type 
eigen-states has much to do with which of the
three operators $\hat{p}^{cons}_x$, $\hat{p}^{cons}_y$ or
$\hat{L}^{cons}_z$ is diagonalized simultaneously with the Landau 
Hamiltonian, but it has little to do with the choices of gauge in the 
Landau problem.
We have no objection to the first half part of the above statement,
but the problem is the last half part. In our opinion, 
whether it has really nothing to do with the choice of gauge in the 
Landau problem is a highly nontrivial question, which concerns 
the deep meaning of the {\it gauge symmetry concept} in physics.
Importantly, the answer to this question is inseparably connected with
observability or nonobservability question of these three quantities. 

A remarkable feature of the above three conserved operators is 
that they transform {\it covariantly} under an arbitrary gauge 
transformation just like the mechanical momentum operators 
$\hat{p}^{mech}_x$ and $\hat{p}^{mech}_y$ and the mechanical 
OAM operator $\hat{L}^{mech}_z$, which are widely believed to 
correspond to {\it genuinely} gauge-invariant observable quantities.
%(As we shall discuss in the paper, this feature is the very
%origin of confusion, which needs careful consideration.) 
On the basis of this fact together with the belief that there is only
one gauge class in the Landau problem,
the author of \cite{Govaerts2023} claims that the three conserved 
quantities also correspond to observables at least in principle. 
In the present paper, we will show that
this claim is incompatible with the celebrated
{\it gauge principle}, regardless of a  conspicuous 
disparity or difference between the two theoretical formulations of 
the Landau problem, i.e. the traditional formulation \cite{Landau1930,
Landau-Lifschitz, Vallentine1998, VGW2006, Goerbig2009, Tong2016} 
and the gauge-potential-independent formulation 
\cite{Konstantinou2016, Konstantinou2017, Haugset1993, GHM2009,
WKZ2018, WH2022, Govaerts2023}. 

\vspace{1mm}
The paper is organized as follows. In sect.\ref{sec2}, on the basis of the 
traditional treatment of the Landau problem, we recall readers of 
general features of the Landau eigen-functions in different gauge choices.
In particular,  by paying careful attention to the role of the {\it quantum 
guiding center} (or orbit center) concept in the Landau problem
\cite{JL1949, van_Enk2020, KWZZ2020}, 
we show how and why the probability and current 
distributions of the Landau electron are so different in different gauges.
Next, in sect.\ref{sec3}, we briefly review the essence of the 
gauge-potential-independent formulation of the Landau problem
with a particular intention of explaining what is the controversy
over observability or nonobservability of 
the quantum numbers $m$ and $k_x$ characterizing 
the two types of Landau eigen-states.
In sect.\ref{sec4}, we try to convince non-observable nature of the quantum number 
$k_x$ characterizing the Landau eigen-states in the 1st Landau gauge
with the help of a concrete example, i.e. the familiar quantum
Hall effect. Next, in sect.\ref{sec5}, by paying attention to the similarity
and dissimilarity between the Landau problem and the problem of
the 2-dimensional Harmonic oscillator, we demonstrate that 
the quantum number $m$ characterizing the Landau eigen-states 
in the symmetric gauge does not correspond to direct observables.
In sect.\ref{sec6}, we briefly discuss implication of the knowledge gained in the 
present paper on the so-called gauge-invariant decomposition problem
of the nucleon spin.
Finally, in sect.\ref{sec7}, we summarize what we have learned from the Landau problem
about easily misunderstandable physical meaning of gauge-invariance issue
in quantum mechanics.

\section{General features of the Landau eigen-states in different gauges}
\label{sec2}

In classical mechanics, the electron in a uniform magnetic field makes
a cyclotron motion around some fixed point $(X, Y)$ in the
2-dimensional plane.
This center of cyclotron motion is sometimes called 
the {\it guiding center} or simply the {\it orbit center}.
Obviously, it is a constant of motion in classical mechanics.
When going to quantum mechanics, the guiding center coordinates 
become quantum operators expressed with the position 
operators $(\hat{x}, \hat{y})$ and the velocity 
operators $(\hat{v}_x, \hat{v}_y)$ as
\begin{equation}
 \hat{X} \ = \ \hat{x} \ - \ \frac{\hat{v}_y}{\omega_c}, \ \ \ \ \ 
 \hat{Y} \ = \ \hat{y} \ + \ \frac{\hat{v}_x}{\omega_c} .
\end{equation}
Here, $\omega_c$ is the cyclotron frequency given by 
$\omega_c = \frac{e \,B}{m_e}$ with $B$ being  the strength of the 
uniform magnetic field and with $m_e$ and $- \,e \, (e > 0)$ being
the mass and the charge of the electron, respectively. 
Note that the velocity operator
is related to the mechanical (or kinetic) momentum operator $\hat{\bm{p}}_{mech}$
of the electron through the relation $\hat{\bm{v}} = \hat{\bm{p}}^{mech} / m_e$.
(In the following, assuming that no confusion arises, we shall omit 
hat symbol indicating quantum operators, for notational simplicity.)
Interestingly, the guiding center coordinates are constants of motion
also in quantum mechanics in the sense that they commute with
the Hamiltonian $H$ of the Landau system \cite{Vallentine1998, JL1949} : 
\begin{equation}
 [ X, H ] \ = \ [ Y, H ] \ = \ 0 ,
\end{equation}
where $H$ is given as
\begin{equation}
 H \ = \ \frac{1}{2 \,m_e} \,\left( \bm{p} \ + \ e \,\bm{A} \right)^2.
\end{equation}
Here, the vector potential $\bm{A}$ is supposed to reproduce
the uniform magnetic field by the relation $\nabla \times \bm{A} = B\,\bm{e}_z$.
Very importantly, however, $X$ and $Y$ do not commute with each other.
Rather, they satisfy the following commutation relation : 
\begin{equation}
 [ X, Y ] \ = \ i \,l^2_B ,
\end{equation}
where $l_B = 1 \,/ \,\sqrt{e \,B}$ is called the magnetic length in the
Landau system. 
This non-commutability
of $X$ and $Y$ makes physical interpretation of the guiding center
coordinates in quantum mechanics far less intuitive as compared with 
the classical case \cite{Vallentine1998}.
The purpose of the present section is to reveal mysterious nature
of the quantum mechanical Landau eigen-functions by paying
maximum attention to the role of {\it quantum guiding center}.

Let us start the discussion with the familiar eigen-functions of the 
Landau Hamiltonian in the {\it symmetric gauge}. 
With the choice of the symmetric gauge
potential $\bm{A} = \frac{1}{2} \,B \,( - \,y, x)$, 
%$\bm{A} = \frac{1}{2} \,B \,( - \,y \,\bm{e}_x + x \,\bm{e}_y )$,
the Landau Hamiltonian takes the following form : 
\begin{eqnarray}
 H &=& \frac{1}{2 \,m_e} ( p^2_x \,+ \,p^2_y ) \ + \ 
 \frac{( e \,B )^2}{8 \,m_e} \,( x^2 \,+ \,y^2 ) \ + \ 
 \frac{e \,B}{2 \, m_e} \, ( x \,p_y \,- \,y \,p_x ) \nonumber \\
 &=& \frac{1}{2 \,m_e} ( p^2_x \,+ \,p^2_y )  \ + \   
 \frac{1}{2} \,m_e \,\omega^2_L \,r^2  \ + \  
 \omega_L \,L^{can}_z . \hspace{5mm}
\end{eqnarray}
Here, $\omega_L = (e \,B) / (2 \,m_e) = \omega_c / 2$ is called the 
Larmor frequency, while $L^{can}_z \equiv x \,p_y - y \,p_x$
is the ordinary canonical orbital angular angular momentum (OAM) operator.
This form of Hamiltonian has a {\it rotational symmetry}
around the $z$-axis.
The eigen-functions of the above Hamiltonian are well-known and
given in the form
\begin{equation}
 \Psi^{(S)}_{n,m} (x, y) \ = \ \frac{e^{\,i \,m \,\phi}}{\sqrt{2 \,\pi}} \,\,
 R_{n,m} (r),
\end{equation}
where $R_{n,m} (r)$ denotes the radial wave function given by
\begin{eqnarray}
 R_{n,m} (r) &=& 
 N_{n,m} \,\,\left( \frac{r^2}{2 \,l^2_B} \right)^{\vert m \vert \,/\,2} \,\,
 r^{ - \,\frac{r^2}{4 \,l^2_B}}  \,\,
 L^{\vert m \vert}_{n - \frac{\vert m \vert + m}{2}}
 \,\left( \frac{r^2}{2 \,l^2_B} \right) , \hspace{5mm}
\end{eqnarray}
with the use of the associated Laguerre polynomial $L^\alpha_k (x)$.
Here, $n$ is the familiar Landau quantum number taking non-negative
integer, while $m$ is an integer subject to the constraint $m \leq n$.
To be more explicit, $\Psi^{(S)}_{n, m} (x, y)$ are the simultaneous eigen-functions 
of the Landau Hamiltonian and the canonical OAM
operator $L^{can}_z = - \,i \,\left( x \,\frac{\partial}{\partial y} - 
y \,\frac{\partial}{\partial x} \right)= - \,i \,\frac{\partial}{\partial \phi}$ :
\begin{eqnarray}
 H \,\Psi^{(S)}_{n, m} (x, y) \, &=& \left( 2 \,n \,+ \,1 \right) \,\omega_L \,
 \Psi^{(S)}_{n, m} (x, y), \hspace{5mm} \\
 L^{can}_z \,\Psi^{(S)}_{n, m} (x, y) &=& m \,\Psi^{(S)}_{n, m} (x, y) .
\end{eqnarray}
We also recall that the quantum number 
$n - \frac{\vert m \vert + m}{2} \equiv n_r$ appearing in the above
eigen-functions $\Psi^{(S)}_{n,m} (x,y)$ represents the number of nodes 
in the radial wave function $R_{n, m} (r)$. 

\vspace{1mm}
Although it is not necessarily very popular, the guiding center coordinates
are quite important quantities in controlling the behaviors of
the Landau wave functions and the associated probability and current
densities of the Landau electron \cite{Vallentine1998}.
Especially important here are the following two 
quantities \cite{JL1949},\cite{WKZZ2020}.
One is the square of the cyclotron radius (operator) defined by
\begin{equation}
 r^2_c \ \equiv \ ( x \,- \,X )^2 \ + \ (y \,- \,Y )^2 .
\end{equation}
Another is the square of the distance between the guiding center
and the coordinate origin given by
\begin{equation}
 R^2 \ \equiv \ X^2 \ + \ Y^2 .
\end{equation}
As shown by Johnson and Lippmann many years ago \cite{JL1949} , 
there is a remarkable relationship between these 
two quantities and the canonical OAM operator
$L^{can}_z$, which is given as
\begin{equation}
 L^{can}_z \ = \ \frac{1}{2 \,l^2_B} \,\left( r^2_c \ - \ R^2 \right) .
\end{equation}
For convenience, we introduce the notation $\langle O \rangle$
to denote the expectation value of any operator $O$ between
the Landau eigen-functions in the symmetric gauge, i.e. 
\begin{eqnarray}
 \langle O \rangle \equiv \int\!\!\!\int  dx \,dy \, 
 \Psi^{(S) *}_{n, m} (x, y) \,O \,
 \Psi^{(S)}_{n, m} (x, y) . \hspace{2mm}
\end{eqnarray}
The expectation values of the operators $r^2_c$ and $R^2$
can easily be evaluated as \cite{JL1949, WKZZ2020}
\begin{eqnarray}
 \langle r^2_c \rangle &=& ( 2 \,n \ + \ 1 ) \,l^2_B , \label{rc2} \\
 \langle R^2 \rangle &=& ( 2 \,n \ - \ 2 \,m \ + \ 1 ) \,l^2_B .
 \label{R2}
\end{eqnarray}
Note that these answers are naturally consistent with the relation
\begin{equation}
 \langle L^{can}_z \rangle \ = \ m
\end{equation}
The above relation (\ref{R2}) plays a particularly important role 
for understanding the physical meaning of the quantum number $m$.
For a fixed value of the Landau quantum number $n$, the quantum
number $m$ has little to do with the rotational motion, but instead 
it is related to the radial position of the guiding center with respect
to the coordinate origin \cite{Vallentine1998}.
In particular, from Eqs.(\ref{rc2}) and (\ref{R2}), one immediately
notice that the following remarkable relations hold \cite{WKZZ2020} : 
\begin{eqnarray}
 \sqrt{\langle r^2_c \rangle} \ & >& \ \sqrt{\langle R^2 \rangle} \ \ \ 
 \mbox{when} \ \ \ m > 0, \\
 \sqrt{\langle r^2_c \rangle} \ &=& \ \sqrt{\langle R^2 \rangle} \ \ \ 
 \mbox{when} \ \ \ m = 0, \\
 \sqrt{\langle r^2_c \rangle} \ &<& \ \sqrt{\langle R^2 \rangle} \ \ \ 
 \mbox{when} \ \ \ m < 0 . 
\end{eqnarray}
This shows that the sign of the magnetic quantum number 
$m$ is intimately connected with the magnitude correlation
between $r_c$ and $R$ \cite{WKZZ2020}. 

%%%%%%%%%
%%%%%%%%%
\vspace{4mm}
\begin{figure}[h]
%\begin{center}
\centering
\includegraphics[width=10.0cm]{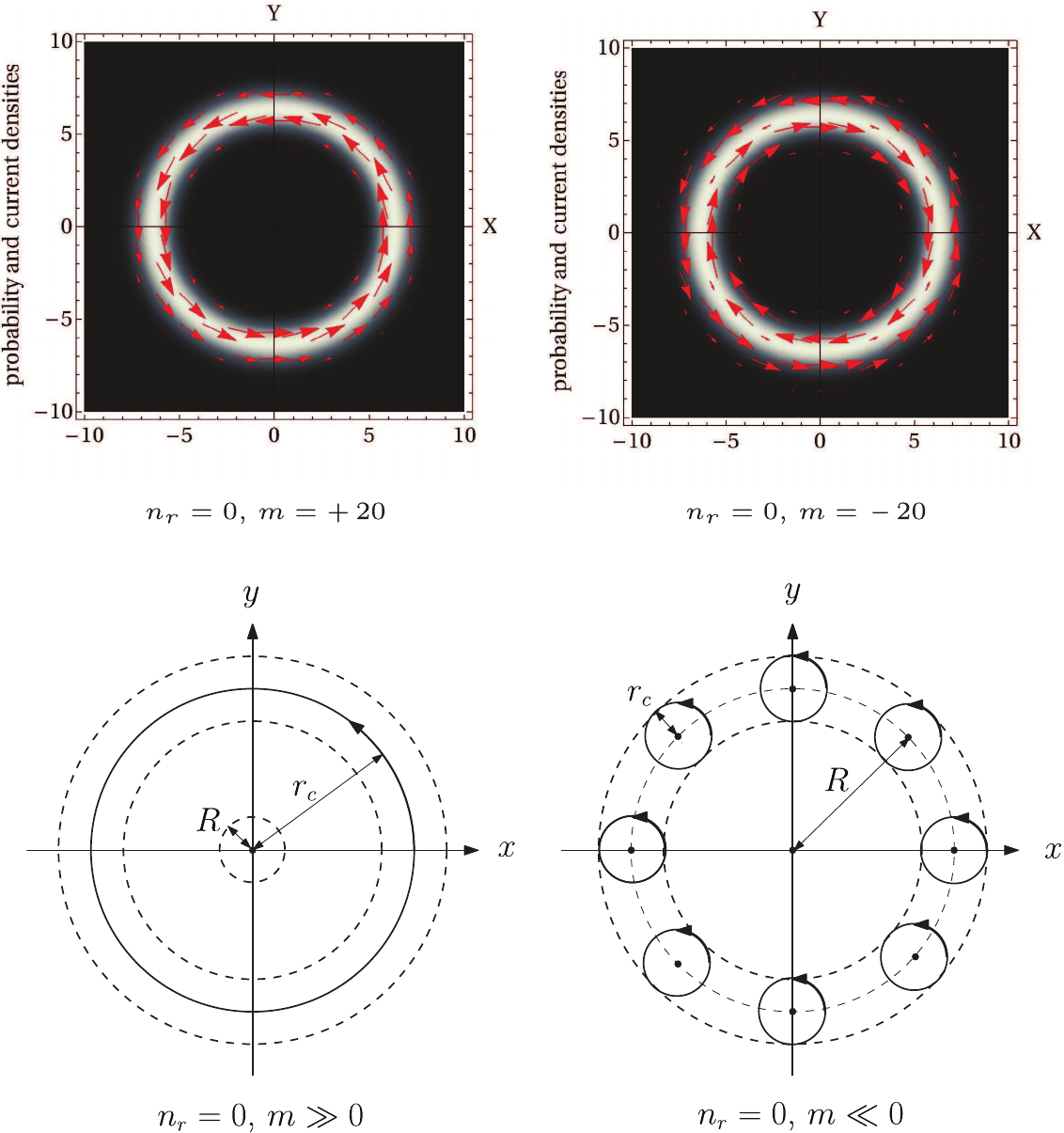}
\caption{Two figures on the upper panel represent the probability
distributions (gray scale) and the current distributions (red arrows
in color) of the Landau electron in the symmetric gauge. 
The left figure corresponds to the
Landau state with $n_r = 0, m = + \,20$, while the right figure
corresponds to the state with $n_r = 0, m = - \,20$.
Shown on the lower panel are the schematic pictures of the
quantum mechanical cyclotron motion of the Landau electron.
The left figure corresponds to the state with $n_r = 0$ and
$m \gg 0$, while the right figure corresponds to the state
with $n_r = 0$ and $m \ll 0$. In both figures,
$r_c$ represents the cyclotron radius, while $R$ does the
distance between the guiding center and the coordinate origin.
Note that the position of the quantum guiding center is statistically
distributed on the circle of radius $R$ with equal probability.}
%\end{center}
\end{figure}
%%%%%%%%%
%%%%%%%%%

For help in more visual understanding,
we show on the upper panel of Fig.1 the probability and current
densities of the electron corresponding to the two Landau
eigen-states with $n_r = 0, m = + \,20$ (or $n = 20, m = + \,20$)
and with $n_r = 0, m = - \,20$ (or $n = 0, m = - \,20$).
In these two figures, higher probability density regions are indicated by
brighter (white) color, whereas lower probability regions are
drawn by darker (black) color.
Note that the eigen-energies of the above two states are significantly
different. In fact, the eigen-energy of the state with $n = 20, m = + \,20$
is $( 2 \times 20 + 1) \,\omega_L = 41 \,\omega_L$, while that of the
state with $n = 0, m = \,- \,20$ is $( 2 \times 0 + 1) \,\omega_L
= \omega_L$.
Nonetheless, one can see that the probability densities of these
two states perfectly coincide with each other. 
This is due to the following
nontrivial relation for the radial wave function as pointed out
in \cite{WKZZ2020} : 
\begin{equation}
 R_{n - m, - \,m} (r) \ = \ R_{n, m} (r) .
\end{equation}
Interestingly, however, the electron current densities shown by
red arrows are drastically different for these two states.
In fact, for the state with $n_r = 0, m = + \,20$, the current
is flowing in a counter-clock-wise direction in both of
the inner and outer regions of the high probability region.
On the other hand, for the state with $n_r = 0, m = - \,20$,
the current is flowing in a counter-clock-wise direction in the
outer region, while it is flowing in clock-wise direction in the
inner part of the high probability region. 

\vspace{1mm}
The reason of this critical difference can be understood from
the schematic pictures  of the quantum mechanical cyclotron 
motion of the Landau electron shown on the lower panel of Fig.1.
The left figure here corresponds to the state with $n_r = 0, m \gg 0$,
while the right figure to the state with $n_r = 0, m \ll 0$.
For the first state with $n_r = 0$  and
$m$ being positive integer with large absolute value,
the cyclotron radius $r_c$ is much larger than the distance $R$ between
the orbit center and the coordinate origin, so that the electron
is circulating  in the counter-clock-wise
direction in the whole high probability region schematically shown
by the two (dotted) concentric circles.
On the other hand, for the second state with $n_r = 0$ and $m$
being negative integer with large absolute value,
the cyclotron radius $r_c$ is much smaller than the distance $R$
between the orbit center and the coordinate origin.
Here, highly non-intuitive in quantum mechanics is that there is
an inherent uncertainty in the position of the orbit center
because of the non-commutativity of the guiding center coordinates,
i.e. $[ X, Y] \neq 0$. The rotational symmetry of the problem dictates
that that the guiding center in
quantum mechanics is distributed on the circle of radius $R$ with
equal probability. Roughly speaking, it means that the electron is 
making a cyclotron motion around arbitrary points on the circle with
radius $R$. This explains the reason why the flow of the current
is counter-clock-wise in the outer region of the high probability region,
while the flow of the current is clock-wise in the inner part \cite{WKZZ2020}.
(See the right picture on the lower panel of Fig.1.)

\vspace{2mm}  
Next, with the choice of the {\it 1st Landau gauge} potential 
$\bm{A} = B \,( - \,y, 0)$, the Landau Hamiltonian takes the following
form : 
\begin{equation}
 H \ =\ \frac{1}{2 \,m_e} \,( p^2_x \ + \ p^2_y ) \ - \ 2 \,\omega_L \,y \,p_x
 \ + \ \frac{1}{2} \,\omega^2_L \,y^2 .
\end{equation}
This form of Landau Hamiltonian shows the {\it translational symmetry}
with respect to the $x$-direction. In fact, since $H$ does not depend on 
the coordinate $x$, its eigen-functions are obtained in the following form : 
\begin{equation}
 \Psi^{(L_1)}_{n, k_x} (x, y) \ = \ \frac{e^{\,i \,k_x \,x}}{\sqrt{2 \,\pi}} \,\,
 Y_n (y) ,  \label{L_1gauge-wf}
\end{equation}
where
\begin{equation}
 Y_n (y) \ = \ N_n \,\,e^{\,- \, \frac{(y - y_0)^2}{2 \,l^2_B}} \,
 H_n \left( \frac{y - y_0}{l_B} \right) , 
\end{equation}
with $ y_0 \, = \, l^2_B \,k_x$.
%
%\begin{equation}
% y_0 \ = \ l^2_B \,k_x .
%\end{equation}
%
They are the simultaneous eigen-functions of the Landau Hamiltonian $H$
and the canonical momentum operator $p^{can}_x = - \,i \,\frac{\partial}{\partial x}$ : 
\begin{eqnarray}
 H \,\Psi^{(L_1)}_{n, k_x} (x, y) \, &=& ( 2 \, n \,+ \,1 ) \,
 \omega_L \,\Psi^{(L_1)}_{n, k_x} (x, y), \hspace{6mm} \\
 p^{can}_x \,\Psi^{(L_1)}_{n, k_x} (x, y) &=& k_x \,\Psi^{(L_1)}_{n, k_x} (x, y) .
 \label{L_1gauge_px-wf}
\end{eqnarray}
Physical interpretation of the above eigen-states is again far from
intuitive. Classically, one expects that the Landau electron makes a 
circular motion around some fixed point.
However, it is not so easy to read such a behavior from the above
eigen-functions. The $x$-dependence of (\ref{L_1gauge-wf}) takes a form of
plane-wave, while the $y$-dependence of it shows a localized
Harmonic oscillator around $y = y_0$. 
The concept of guiding center plays an important role
also in understanding the behavior of these eigen-functions in the 1st 
Landau gauge. First, the relation $y_0 = l^2_B \,k_x$ indicates that
the eigen-equation (\ref{L_1gauge_px-wf}) can also be regarded as an eigen equation
for the guiding center coordinate (operator) $Y$ expressed as \cite{Vallentine1998} 
\begin{equation}
 Y \,\Psi^{(L_1)}_{n, y_0} (x, y) \ = \ y_0 \,\Psi^{(L_1)}_{n, y_0} (x, y) . 
 \label{L_1gauge_Y-wf}
\end{equation}
The validity of this conjecture can easily be confirmed as follows.
With the choice of the 1st Landau gauge potential
$\bm{A}^{(L_1)} = B \,( - \,y, 0)$, the mechanical momentum
operator $p^{mech}_x$ takes the following form : 
\begin{equation}
 p^{mech}_x \ \equiv \ p^{can}_x \ + \ e \,A_x \ = \ 
 p^{can}_x \ - \ e \,B \,y .
\end{equation}
Then, the guiding center coordinate $Y$ defined as 
$Y = y \,+ v_x \,/\,\omega_c = y \,+ \,p^{mech}_x \,/\,
(e \, B)$ reduces to
\begin{eqnarray}
 Y &=& y \ + \,\frac{1}{e \,B} \,( p^{can}_x - e \,B \,y ) 
 \ = \  \frac{1}{e \,B} \,\,p^{can}_x \ = \ l^2_B \, \,p^{can}_x ,
 \hspace{4mm}
\end{eqnarray}
which means that the guiding center coordinate $Y$ and
the canonical momentum $p^{can}_x$ basically represent the same
entity aside from a proportionality constant. 
The equation (\ref{L_1gauge_Y-wf}) then means
that $y_0$ in the Landau eigen-functions $\Psi^{(L_1)}_{n, k_x} (x, y)$ 
represents the $y$-coordinate of the guiding center.
However, the price to pay for it is that the $x$-coordinate of
the guiding center becomes totally uncertain, that is, it is 
distributed uniformly along the line $y = y_0$. 

\vspace{2mm}
To visually understand such circumstances, we show on the upper 
panel of Fig.2 the probability and
current densities of two Landau states.
The left figure corresponds to the state with $n = 0$ and $y_0 = + \,5$,
whereas the right figure to the state with $n=0$ and $y_0 = - \,5$.  
One clearly sees that the high probability region for the former state
is distributed along the line $y = + \,5$, while the high probability
region for the latter state is distributed along the line $y = - \,5$.
Also interesting is the behavior of the current densities shown by red
arrows. In both cases of $y_0 = + \,5$ and $y_0 = - \,5$, the current
is flowing to the left in the upper (larger $y$) part of the high
probability region, whereas it is flowing to the right in the
lower (smaller $y$) part of the high probability region.
These behaviors can easily be understood from the two schematic
pictures illustrated on the lower panel of Fig.2.
These figures show that the electron is rotating counter-clock-wise
around the guiding center, which is uniformly distributed along the
line $y = y_0 = + \,5$ or $y = y_0 = - \,5$. It is clear that this explains 
the reason why the  current is flowing to the left in the upper part of the
higher probability band, while it is flowing to the right in the lower
part  of the higher probability band.

%\vspace{-2mm}
\begin{figure}[h]
%\begin{center}
\centering
\includegraphics[width=10.0cm]{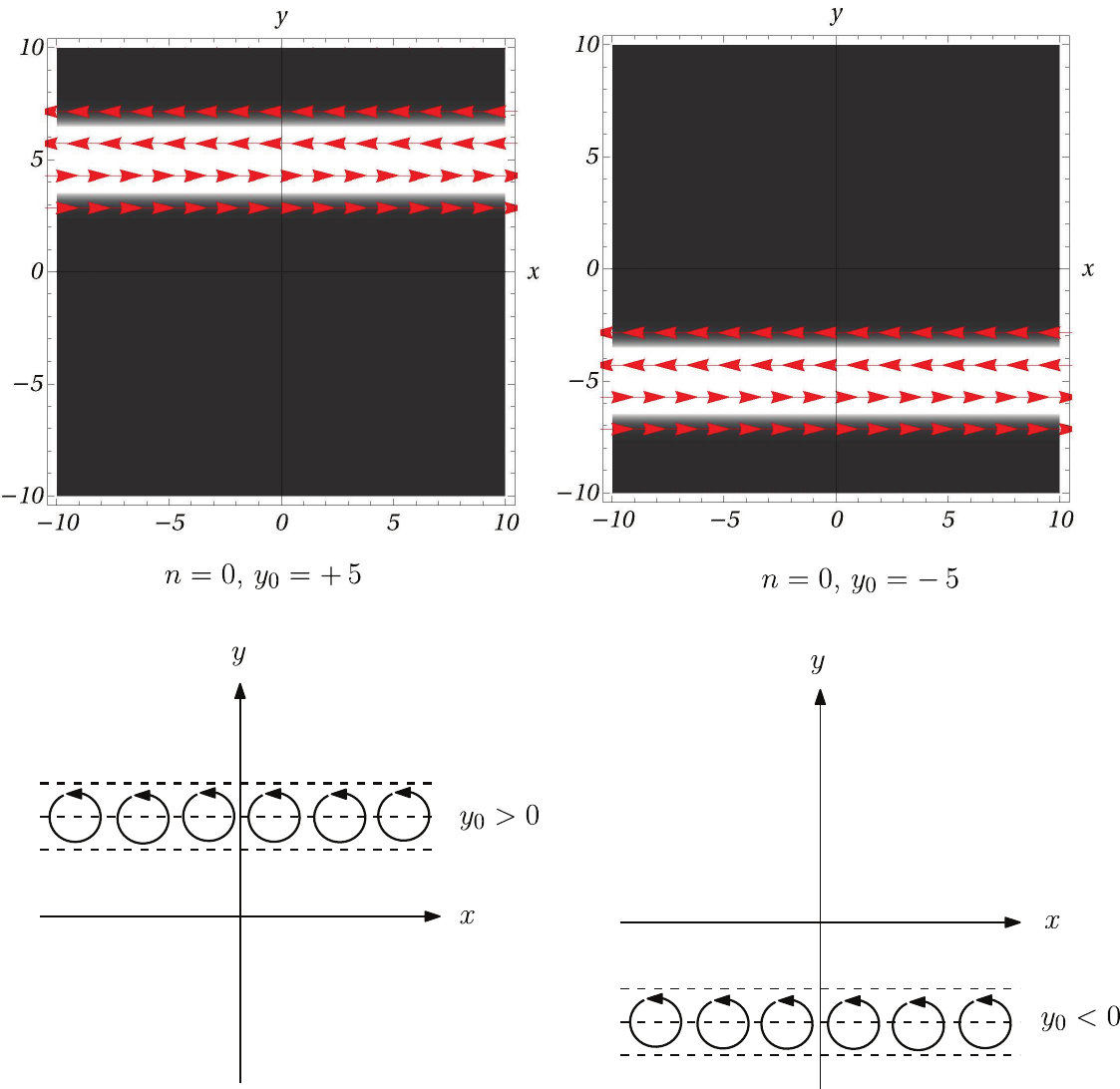}
\vspace{0mm}
\caption{Two figures on the upper panel represent the probability
distributions (gray scale) and the current distributions (red arrows
in color) of the Landau electron in the 1st Landau gauge. 
The left figure corresponds to the
Landau state with $n = 0, y_0 = + \,5$, while the right figure
corresponds to the state with $n = 0, y_0 = - \,5$.
Shown on the lower panel are the schematic pictures of the
quantum mechanical cyclotron motion of the Landau electron.
The left figure corresponds to the state with $n = 0, y_0 = + \,5$,
while the right figure corresponds to the state with $n = 0, y_0 = - \,5$.
Note that the positions of the guiding center are respectively
distributed on the lines with $y = y_0 = + \,5$ and $y = y_0 = - \,5$
with equal statistical probability.}
%\end{center}
\end{figure}

\vspace{1mm}
In any case, we  now understand that the quantum mechanical 
electron's probability and the current distributions of the Landau
eigen-states take remarkably different forms in different gauges,
i.e. the symmetric gauge and the 1st Landau gauge.
Still, if one believes the celebrated {\it gauge principle}, one would expect
that physical observables are independent of these choices of 
gauge\footnote{Note however that the gauge principle never 
demands that the converse is always true.}.
As we shall discuss in the following, to explicitly confirm this fact turns out
to be far more delicate problem than is naively anticipated.

%%%%%%%%%%%%%%%%%%%%%%%%%%%%%%%%%%%%%%%%%%
%%%%%%%%%%%%%%%%%%%%%%%%%%%%%%%%%%%%%%%%%%
\section{Controversy over observability of 
the quantum numbers $m$ and $k_x$ characterizing 
the two types of Landau eigen-states}
\label{sec3}

In the traditional interpretation of the gauge choice in the Landau problem
which we follow in the previous section,
the quantum number $k_x$ is the eigen-value of the canonical momentum
operator $\hat{p}^{can}_x$ in the 1st Landau gauge eigen-states, whereas the quantum
number $m$ is the eigen-value of the canonical OAM operator $L^{can}_z$ 
in the symmetric gauge eigen-states.
Note that the canonical momentum as well as the canonical OAM are
standardly believed to be {\it gauge-variant} quantities. 
Then, if one believes the celebrated gauge principle, neither $k_x$ nor $m$
would correspond to {\it direct} experimental observables.
However, this widely-accepted view has been suspected in a recent paper by
Govaerts \cite{Govaerts2023} based on logical development  of the 
gauge-potential-independent formulation of the Landau problem. 
The grounds of this claim can be traced back to the observation
that the quantum numbers $k_x$ and $m$ are related to the conserved 
Noether charges of the Landau Hamiltonian, the existence of which is 
guaranteed totally independently of the choice of the {\it gauge potential configuration}.
On the basis of this not-so-familiar knowledge together with 
the belief that there is {\it only one gauge class} in the Landau problem
within the gauge-potential-independent formulation,
it was claimed that $k_x$ and $m$ both correspond to 
physical (or observable) quantities at least in principle. 

\vspace{1mm}
To explain in more detail about the controversy over the physical 
significance of the quantum numbers $k_x$ and $m$ appearing in the two different 
types of Landau eigen-states, we need to briefly remind the essence of the
gauge-potential-independent formulation of the Landau problem, which 
was independently  proposed by several authors in different contexts as well as in
slightly different forms \cite{Haugset1993, GHM2009, Konstantinou2016,
Konstantinou2017, WKZ2018, WH2022}.
To make the discussion as elementary as possible, we think it
most comprehensive to explain the problem based on the 
theoretical formulation developed in \cite{WH2022}, which is based only on
the elementary knowledge of quantum mechanics supplemented with 
the basics of gauge transformation theory.
First, it is very important to reconfirm the fact that the 
Landau Hamiltonian explicitly depends on the {\it gauge-dependent} vector 
potential $\bm{A}$ as
\begin{equation}
 \hat{H} (\bm{A}) \ = \ \frac{1}{2 \,m_e} \,
 \left( - \,i \,\nabla \ + \ e \,\bm{A} \right)^2 ,
\end{equation}
where $\bm{A}$ is an arbitrary gauge potential, which reproduces the
uniform magnetic field through the relation $\nabla \times \bm{A} = B \,\bm{e}_z$.

\vspace{1mm}
A nontrivial observation by several authors \cite{Konstantinou2016,
Konstantinou2017, WKZ2018, WH2022, Govaerts2023}
is that there exist the following three {\it conserved quantities} in the Landau problem
independently of the choice of the gauge potential $\bm{A}$.
They are two conserved momenta and one conserved OAM given 
by\footnote{They are the same quantities as those called the {\it pseudo momenta} 
and the {\it pseudo OAM} in some literature \cite{Yoshioka2002, Konstantinou2016,
Konstantinou2017, KWZZ2020}.}
\begin{eqnarray}
 \hat{p}^{cons}_x (\bm{A}) &\equiv& - \,i \,\frac{\partial}{\partial x} \ + \ 
 e \,A_x \ + \ e \,B \,y  \ = \ 
 \hat{p}^{mech}_x (\bm{A}) \ + \ e \,B \, y , \label{def_px_cons} \\
 \hat{p}^{cons}_y (\bm{A}) &\equiv& - \,i \,\frac{\partial}{\partial y} \ + \ 
 e \,A_y \ - \ e \,B \,x \ =\  
 \hat{p}^{mech}_x (\bm{A}) \ - \ e \,B \, x , \label{def_py_cons} \\
 \hat{L}^{cons}_z (\bm{A}) &\equiv& - \,i \,\frac{\partial}{\partial \phi} \ + \ 
 e \,r \,A_\phi \ - \ \frac{1}{2} \,e \,B \,r^2 \ = \  
 \hat{L}^{mech}_z (\bm{A}) \ - \ \frac{1}{2} \,e \,B \,r^2 . \hspace{5mm}
 \label{def_Lz_cons}
\end{eqnarray}
In these equations, $\hat{p}^{mech}_x (\bm{A})$ and $\hat{p}^{mech}_x (\bm{A})$
respectively denote the $x$- and $y$-components of the familiar mechanical 
momentum operators, while $\hat{L}^{mech}_z (\bm{A})$ does the $z$-component 
of the mechanical OAM operator.
In the following argument, since the operator $\hat{p}^{cons}_y (\bm{A})$
can be treated just in the same manner as $\hat{p}^{cons}_x (\bm{A})$,
we concentrate on the discussion of the two operators
$\hat{p}^{cons}_x (\bm{A})$ and $\hat{L}^{cons}_z (\bm{A})$.
Remarkably, the conservation of these quantities holds independently of the
choice of the vector potential \cite{Konstantinou2016,Konstantinou2017,WH2022}. 
In fact, it can be shown that they commute 
with the Landau Hamiltonian for {\it arbitrary choice} of the gauge potential 
$\bm{A}$, i.e.
\begin{equation}
 [ \hat{p}^{cons}_x (\bm{A}), \, \hat{H} (\bm{A}) ] \ = \ 0, \ \ \ \ 
 [ \hat{L}^{cons}_z (\bm{A}), \, \hat{H} (\bm{A}) ] \ = \ 0 .
\end{equation}
Very importantly, however, they do not commute with each 
other \cite{Haugset1993, WH2022},
\begin{equation}
 [ \hat{p}^{cons}_x (\bm{A}), \,\hat{L}^{cons}_z (\bm{A}) ] \ \neq \ 0.
\end{equation}
This means that these two operators cannot be simultaneously diagonalized
together with the Landau Hamiltonian $\hat{H} (\bm{A})$.
One is forced to select either of the following two options : 

\vspace{2mm}
\begin{itemize}
 \item[(1)] diagonalize $\hat{L}^{cons}_z (\bm{A})$ and $\hat{H} (\bm{A})$, 
 simultaneously.
 \vspace{2mm}
 \item[(2)] diagonalize $\hat{p}^{cons}_x (\bm{A})$ and $\hat{H} (\bm{A})$,
 simultaneously. 
\end{itemize}

\vspace{2mm}
\noindent
The easiest way to realize the 1st option is to start with the Landau
eigen-states $\vert \Psi^{(S)}_{n, m} \rangle$ in the symmetric gauge, 
which are known to satisfy the simultaneous eigen-equations as follows : 
\begin{eqnarray}
 \hat{H} (\bm{A}^{(S)}) \,\vert \Psi^{(S)}_{n, m} \rangle &=& 
 (2 \,n + 1) \,\omega_L \,\vert \Psi^{(S)}_{n, m} \rangle , \label{Eigen_HS} 
 \hspace{4mm} \\
 \hat{L}^{can}_z \,\vert \Psi^{(S)}_{n, m} \rangle &=& 
 \ m \,\vert \Psi^{(S)}_{n, m} \rangle . \label{Eigen_Lcan}
\end{eqnarray}
Here, $\bm{A}^{(S)} = \frac{1}{2} \,B \,( - \,y, \,x)$ represents the vector 
potential in the symmetric gauge, 
whereas $\hat{L}^{can}_z = - \,i \,\frac{\partial}{\partial \phi}$ is the 
standard canonical OAM operator.
An important observation here is that the eigen-equation (\ref{Eigen_Lcan}) 
can also be expressed in the following form,
\begin{equation}
 \hat{L}^{cons}_z (\bm{A}^{(S)}) \,\vert \Psi^{(S)}_{n, m} \rangle  = \ 
 m \,\vert \Psi^{(S)}_{n, m} \rangle . \label{Eigen_Lcons}
\end{equation}
This is because $\hat{L}^{cons}_z (\bm{A})$ reduces to the canonical
OAM operator when $\bm{A}$ becomes $\bm{A}^{(S)}$,
\begin{equation}
 \hat{L}^{cons}_z (\bm{A} \rightarrow \bm{A}^{(S)} )  \ = \ 
 - \,i \,\frac{\partial}{\partial \phi}
 \ = \ \hat{L}^{can}_z .
\end{equation}
Let us now consider a $U(1)$ gauge transformation
$U^{(\chi)} = e^{\,- \,i \,e \,\chi (x, y)}$, which
transforms the symmetric gauge potential $\bm{A}^{(S)}$ to an arbitrary
gauge potential $\bm{A}^{(\chi)}$. The quantum mechanical representation
of such gauge transformation is represented as
\begin{equation}
 - \,i \,\nabla \ + \ e \,\bm{A}^{(\chi)} \ = \ U^{(\chi)} \,\left(
 - \,i \,\nabla \ + \ e \,\bm{A}^{(S)} \right) \,U^{(\chi)^\dagger} .
\end{equation}
Using this relation, one can verify the
following ({\it gauge-covariant}) transformation properties of the Landau 
Hamiltonian as well as the conserved OAM operator :  
\begin{eqnarray}
 U^{(\chi)} \,\hat{H} (\bm{A}^{(S)}) \,U^{(\chi)^\dagger} 
 &=& \hat{H} (\bm{A}^{(\chi)} ) , \label{GT_HS} \\
 U^{(\chi)} \,\hat{L}^{cons}_z (\bm{A}^{(S)}) \,U^{(\chi)^\dagger} 
 &=& \hat{L}^{cons}_z (\bm{A}^{(\chi)} ) . \label{GT_Lcons} \hspace{4mm}
\end{eqnarray}
After these preparations, suppose that we define new states 
$\vert \Psi^{(\chi)}_{n, m} \rangle$, which are obtained from
the symmetric-gauge eigen-states
$\vert \Psi^{(S)}_{n, m} \rangle$ by operating the $U (1)$ gauge transformation
matrix $U ^{(\chi)}$ as
\begin{equation}
 \vert \Psi^{(\chi)}_{n, m} \rangle \ \equiv \ U^{(\chi)} \,
 \vert \Psi^{(S)}_{n,m} \rangle .
\end{equation}
Now, with the use of Eqs.(\ref{Eigen_HS}) and (\ref{Eigen_Lcons}) together 
with the relation (\ref{GT_HS}) and (\ref{GT_Lcons}), it is straightforward to
show that the following equations hold :  
\begin{eqnarray}
 \hat{H} (\bm{A}^{(\chi)}) \,\vert \Psi^{(\chi)}_{n, m} \rangle &=&
 ( 2 \,n + 1 ) \,\omega_L \,\vert \Psi^{(\chi)}_{n, m} \rangle, \hspace{6mm} \\
 \hat{L}^{cons}_z (\bm{A}^{(\chi)}) \,\vert \Psi^{(\chi)}_{n, m} \rangle &=&
 \ m \,\vert \Psi^{(\chi)}_{n, m} \rangle .
\end{eqnarray}
One therefore finds that the state $\vert \Psi^{(\chi)}_{n,m} \rangle$ are the 
simultaneous eigen-states of the Landau Hamiltonian $\hat{H} (\bm{A}^{(\chi)})$
and the conserved OAM operator $\hat{L}^{cons}_z (\bm{A}^{(\chi)})$
with the eigen-values $( 2 \,n + 1) \,\omega_L$ and $m$, respectively.
In view of the arbitrariness of the choice for the gauge function $\chi$,
this means that there are {\it infinitely many} such states. 

\vspace{1mm}
Similarly, as the 2nd option, one can start with the eigen-states 
$\vert \Psi^{(L_1)}_{n, k_x} \rangle$ in the 1st Landau gauge, which are 
known to satisfy the following simultaneous eigen-equations : 
\begin{eqnarray}
 \hat{H} (\bm{A}^{(L_1)}) \,\vert \Psi^{(L_1)}_{n, k_x} \rangle &=& 
 ( 2 \,n + 1 ) \,\omega_L \,\vert \Psi^{(L_1)}_{n, k_x} \rangle , 
 \label{Eigen_HL1} \hspace{4mm} \\
 \hat{p}^{can}_x \,\vert \Psi^{(L_1)}_{n, k_x} \rangle &=&
 \ \ k_x \,\vert \Psi^{(L_1)}_{n, k_x} \rangle , \label{Eigen_pcan}
\end{eqnarray}
where $\bm{A}^{(L_1)} = B \,( - \,y, \, 0)$ represents the vector potential 
in the 1st Landau gauge, while $\hat{p}^{can}_x = - \,i \,\frac{\partial}{\partial x}$
is the standard canonical momentum operator. 
Since $\hat{p}^{cons}_x (\bm{A})$
reduces to the canonical momentum operator when $\bm{A}$ approaches
$\bm{A}^{(L_1)}$ : 
\begin{equation}
 \hat{p}^{cons}_x (\bm{A} \rightarrow \bm{A}^{(L_1)})  
 \ = \ - \,i \,\frac{\partial}{\partial x}
 \ = \ \hat{p}^{can}_x,
\end{equation}
the eigen-equation (\ref{Eigen_pcan}) can also be expressed in the form
\begin{equation}
 \hat{p}^{cons}_x (\bm{A}^{(L_1)}) \,\vert \Psi^{(L_1)}_{n, k_x} \rangle
 \ = \ k_x \,\vert \Psi^{(L_1)}_{n, k_x} \rangle . \label{Eigen_pcons}
\end{equation}
Then, just as before, let us consider a $U(1)$ gauge transformation
$U^{(\chi^\prime)} (x) = e^{\,- \,i \,e \,\chi^\prime (x, y)}$,
which transforms the 1st Landau gauge potential $\bm{A}^{(L_1)}$ to
an arbitrary gauge potential $\bm{A}^{(\chi^\prime)}$, 
the quantum mechanical representation of which is given as
\begin{equation}
 - \,i \,\nabla \, + \, e \,\bm{A}^{(\chi^\prime)} \, = \, U^{(\chi^\prime)} \,
 \left( - \,i \,\nabla \, + \, e \,\bm{A}^{(L_1)} \right) \,
 U^{(\chi^\prime)^\dagger} .
\end{equation}
Accordingly, one can readily verify that the following relations hold, 
\begin{eqnarray}
 U^{(\chi^\prime)} \,\hat{H} (\bm{A}^{(L_1)}) \,U^{(\chi^\prime)^\dagger} 
 &=& \hat{H} (\bm{A}^{(\chi)} ) , \\
 U^{(\chi^\prime)} \,\hat{p}^{cons}_x (\bm{A}^{(L_1)}) \,U^{(\chi^\prime)^\dagger} 
 &=& \hat{p}^{cons}_x (\bm{A}^{(\chi^\prime)} ) . \hspace{4mm}
\end{eqnarray}
Hence, if one defines the new states $\vert \Psi^{(\chi^\prime)}_{n, k_x} \rangle$
by
\begin{equation}
 \vert \Psi^{(\chi^\prime)}_{n, k_x} \rangle \ = \ U^{(\chi^\prime)} \,
 \vert \,\Psi^{(L_1)}_{n, k_x} \rangle ,
\end{equation}
they clearly satisfy the following eigen-equations : 
\begin{eqnarray}
 \hat{H} (\bm{A}^{(\chi^\prime)}) \,\vert \Psi^{(\chi^\prime)}_{n, k_x} \rangle &=&
 ( 2 \,n + 1 ) \,\omega_L \,\vert \Psi^{(\chi^\prime)}_{n, m} \rangle, \hspace{5mm} \\
 \hat{p}^{cons}_x (\bm{A}^{(\chi^\prime)}) \,\vert \Psi^{(\chi^\prime)}_{n, k_x} \rangle &=&
 \ \ k_x \,\vert \Psi^{(\chi^\prime)}_{n, k_x} \rangle . \label{GT_pcons}
\end{eqnarray}
Again, we find that there are {\it infinitely many} states, which are 
simultaneous eigen-states of $\hat{H} (\bm{A}^{(\chi^\prime)})$ and
$\hat{p}^{cons}_x (\bm{A}^{(\chi^\prime)})$ with the eigen-values
$( 2 \,n + 1) \,\omega_L$ and $k_x$, respectively.

In this way, we are led to concluding that there are infinitely many eigen-states
$\vert \Psi^{(\chi)}_{n, m} \rangle$ with arbitrary gauge potential $\bm{A}^{(\chi)}$, 
which are obtained from the symmetric-gauge eigen-state $\vert \Psi^{(S)}_{n, m} \rangle$ 
by means of $U (1)$ gauge transformations,
and also another class of eigen-states $\vert \Psi^{(\chi^\prime)}_{n, k_x} \rangle$
with arbitrary gauge potential $\bm{A}^{(\chi^\prime)}$, which are obtained from
the 1st Landau-gauge eigen-state $\vert \Psi^{(L_1)}_{n, k_x} \rangle$.
In \cite{WH2022}, these two types of eigen-states are classified into the quantum states 
belonging to {\it different} or {\it inequivalent gauge classes}. 

\vspace{1mm}
This understanding was criticized in a recent paper \cite{Govaerts2023}, however.
According to the viewpoint advocated in that paper, 
the existence of the two classes of eigen-states
$\vert \Psi^{(\chi)}_{n, m} \rangle$ and $\vert \Psi^{(\chi^\prime)}_{n, k_x} \rangle$
has little to do with the choice of gauge, but rather which of the two operators
$\hat{L}^{cons}_z$ or $\hat{p}^{cons}_x$ one wants to have 
diagonalized \cite{Haugset1993}. 
The latter half part of this statement is nothing different from
our own argument explained above. An intricate question here is whether 
the choice of the two types of
eigen-states has really nothing to do with the choice of gauge in the
Landau problem. Here, let us recall the 
crucial statements made in \cite{Govaerts2023}, which deeply concerns the core of
the conflict. They are rephrased as follows by using our notation \cite{WH2022} 
instead of  somewhat complex notation in \cite{Govaerts2023}.

\vspace{1mm}
\begin{itemize}
 \item[(i)] The two bases $\vert \Psi^{(\chi)}_{n, m} \rangle$ 
and $\vert \Psi^{(\chi^\prime)}_{n, k_x} \rangle$ span a same abstract Hilbert space and
they are mutually related through a {\it specific unitary transformation}.
\vspace{1mm}
 \item[(ii)] There is {\it only one gauge class} in the Landau problem.
\vspace{1mm}
 \item[(iii)] Since the conserved quantities like $\hat{p}^{cons}_x$ and 
 $\hat{L}^{cons}_z$  transform {\it gauge-covariantly} just in the same way as the 
 mechanical quantities like  $\hat{p}^{mech}_x$ and $\hat{L}^{mech}_z$, 
 their eigen-values $k_x$ and $m$ in principle correspond to {\it observables}.
\end{itemize}

\vspace{1mm}
First, we point out that the statement (i) is not valid or at least very obscure
in the sense that what is meant by a {\it specific unitary transformation} is not
clearly stated. To show why it is not correct as plainly as possible, let us consider the 
symmetric-gauge eigen-states $\vert \Psi^{(S)}_{n, m} \rangle$
as a representative of $\vert \Psi^{(\chi)}_{n, m} \rangle$, and the 1st
Landau-gauge eigen-state $\vert \Psi^{(L_1)}_{n, k_x} \rangle$ as a representative of 
$\vert \Psi^{(\chi^\prime)}_{n, k_x} \rangle$. It is already known that they are 
connected through the following relation \cite{Haugset1993, WKZ2018, WH2022},
\begin{eqnarray}
 U_0 \,\Psi^{(S)}_{n, m} (x, y)  =  \int d k_x \,U_{n, m} (k_x) \,
 \Psi^{(L_1)}_{n, k_x} (x, y) . \hspace{4mm} \label{superposition}
\end{eqnarray}
In the above equation,  
$U_0 = e^{\,i \,\frac{1}{2} \,e \,B \,x \,y}$ is a $U(1)$ gauge
transformation operator, which transforms the symmetric gauge potential
$\bm{A}^{(S)}$ to the 1st Landau-gauge potential $\bm{A}^{(L_1)}$, while
the quantity $U_{n, m} (k_x)$ is defined as the following matrix element of this
unitary operator $U_0$,
\begin{equation}
 U_{n, m} (k_x) \ \equiv \ \langle \Psi^{(L_1)}_{n, k_x} \,\vert \,U_0 \,\vert \,
 \Psi^{(S)}_{n, m} \rangle.
\end{equation}
Here we omit to show the explicit form of $U_{n,m} (k_x)$, since it is already given 
in \cite{WKZ2018, WH2022, Haugset1993}.
Eq.(\ref{superposition}) above or Eq.(155) in the paper \cite{Govaerts2023} shows 
that the $U(1)$ gauge-transformed states of the symmetric gauge eigen-states
$\Psi^{(S)}_{n, m} (x, y)$ are expressed as a {\it linear combination} of the 
1st Landan-gauge eigen-states $\Psi^{(L_1)}_{n, k_x} (x, y)$.
What is vitally important to recognize here is that it
never means that the eigen-functions $\Psi^{(L_1)}_{n, k_x} (x, y)$ in the 1st Landau
gauge and the eigen-functions $\Psi^{(S)}_{n, m} (x, y)$ in the symmetric gauge 
are connected through a $U(1)$ transformation (or a {\it phase transformation}).
In fact, if these two eigen-states were simply connected through
a phase transformation, the electron's probability and current densities 
corresponding to these two states must be exactly the same, i.e. we would have
\begin{equation}
 \vert \Psi^{(S)}_{n, m} (x, y) \vert^2 \ \stackrel{?}{=} \ 
 \vert \Psi^{(L_1)}_{n, k_x} (x, y) \vert^2.
\end{equation}
Obviously, this is not the case. As we have already shown in sect.\ref{sec2},
the electron probability density $\vert \Psi^{(S)}_{n, m} (x, y) \vert^2$
shows a rotational symmetry around the $z$-axis, while the density
$\vert \Psi^{(L_1)}_{n, k_x} (x, y) \vert^2$ does a translational symmetry
with respect to the $x$-axis, so that they are drastically different
\begin{equation}
 \vert \Psi^{(S)}_{n, m} (x, y) \vert^2 \ \neq \ 
 \vert \Psi^{(L_1)}_{n, k_x} (x, y) \vert^2,
\end{equation}
%

%What is vitally important to recognize here is that the eigen-functions in the 1st Landau
%gauge and those in the symmetric gauge are {\it not} connected through a simple $U(1)$
%transformation (or a {\it phase transformation}).
%Rather, the states $U_0 \,\vert \Psi^{(S)}_{n, m} \rangle$ obtained from the 
%symmetric-gauge eigen-states by means of a $U (1)$ gauge transformation
%$U_0$ are {\it superposition} of the eigen-states $\vert \Psi^{(L_1)}_{n, k_x} \rangle$ 
%in the 1st Landau gauge for the variable $k_x$ with the weight function $U_{n, m} (k_x)$. 
%This is only natural if one remembers the fact that, if the two
%states $\vert \Psi^{(S)}_{n, m} \rangle$ and $\vert \Psi^{(L_1)}_{n, k_x} \rangle$
%{\it were} related through a simple $U (1)$ gauge transformation,
%the electron's probability and current densities corresponding to these two states must
%be exactly the same. On the contrary, however, we have already shown 
%in sect.\ref{sec2} that they are drastically different : 
%
%\begin{equation}
% \vert \Psi^{(S)}_{n, m} (x, y) \vert^2 \ \neq \ 
% \vert \Psi^{(L_1)}_{n, k_x} (x, y) \vert^2,
%\end{equation}
%

\vspace{0mm}    
The above state of affairs should be contrasted with the two classes of eigen-states 
$\vert \Psi^{(\chi)}_{n, m} \rangle$ and $\vert \Psi^{(\chi^\prime)}_{n, k_x} \rangle$
defined by Eqs.(43) and (53). Since either of the gauge transformation
operators in Eq.(43) or Eq.(53) is just a $U (1)$ gauge transformation or a phase 
transformation, it immediately follows that
\begin{equation}
 \vert \Psi^{(\chi)}_{n, m} (x, y) \vert^2 \ = \ 
 \vert \Psi^{(S)}_{n, m} (x, y) \vert^2 ,
\end{equation}
and
\begin{equation}
 \vert \Psi^{(\chi^\prime)}_{n, k_x} (x, y) \vert^2 \ = \ 
 \vert \Psi^{(L_1)}_{n, k_x} (x, y) \vert^2 ,
\end{equation}

\vspace{1mm}
\noindent
for {\it arbitrary choice} of the gauge functions $\chi$ and $\chi^\prime$. 
Naturally, the probability densities of all the 1st class eigen-states
$\vert \Psi^{(\chi)}_{n, m} \rangle$ has a {\it rotational symmetry} around the $z$-axis,
while those of all the 2nd class eigen-states $\vert \Psi^{(\chi^\prime)}_{n, k_x} \rangle$
has a {\it translational symmetry} along the $x$-axis.
The argument above is unquestionably thought to support our understanding that 
they belong to two different $U (1)$ gauge classes {\it even} within the 
gauge-potential-independent formulation of the Landau problem.

\vspace{2mm}
Once not very meaningful property of the claim (i) has been confirmed, 
the fallacy of the claim (ii) would be almost obvious. 
Still, in view of the delicacy of the issue,
we think it useful to give some additional explanation to remove this misunderstanding.
As already explained, the three types of Landau eigen-states are 
respectively characterized by the pair of two quantum numbers $(n, k_x)$, $(n, k_y)$,
or $(n,m)$, and there are infinitely
many states belonging to each of these three families. 
To be more concrete, we showed that any
eigen-states belonging to the 1st family characterized by the pair of
quantum numbers $(n, k_x)$ can be obtained from the eigen-states
in the 1st Landau gauge by performing $U(1)$ gauge 
transformations\footnote{Remember that the fundamental gauge
symmetry of the Landau Hamiltonian is of course abelian $U(1)$ symmetry.}.  
This means that the 1st family of eigen-states all belong to the same
$U (1)$ gauge class in the sense that they are basically {\it identical
except for phases}. Exactly the same can be said for 
the other two families of eigen-states. 
(This state of things is shown in a schematic figure illustrated in Fig.3.)
Obviously, however, the eigen-states belonging to different families
are not connected through simple $U(1)$ gauge transformations
so that we can say that they belong to different or inequivalent 
$U(1)$ gauge class.\footnote{We emphasize that this interpretation plays
a vitally important role when we discuss below the non-observable nature
of conserved quantities (besides the canonical quantities) on the basis
of the gauge principle.} 
%This is clear from the fact that the eigen-states
%belonging to the different families are not connected through $U(1)$ gauge
%ransformations, which confirms that they {\it do not} belong to the same 
%$U(1)$ gauge class. 
These observations clearly reconfirm the validity of our claim 
that these three families of eigen-states in fact belong to {\it different gauge classes}
even in the gauge-potential-independent formulation of the Landau problem.
It obviously contradicts the claim in \cite{Govaerts2023} that there is {\it only one 
gauge class} in this theoretical formulation. 
%%%%%%%%%%%%%%%%%%%%%%%%%%%%%%%%%%%%%%%%%%%%%%

%\vspace{-2mm}
\begin{figure}[h]
%\begin{center}
\centering
\includegraphics[width=11.0cm]{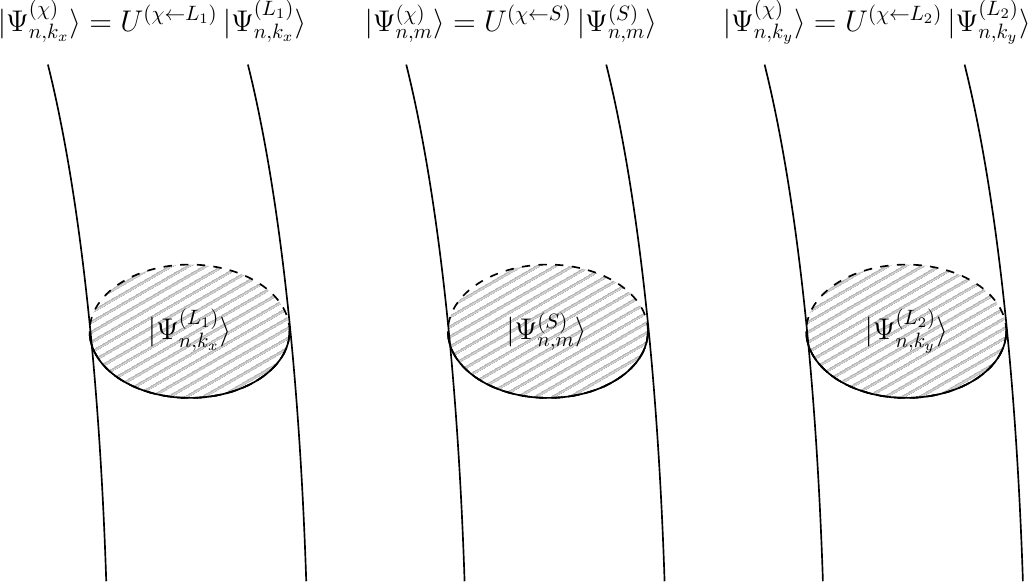}
\vspace{0mm}
\caption{$U(1)$ gauge group structure of the quantum eigen-function
space of the Landau Hamiltonian. 
Here, $\vert \Psi^{(\chi)}_{n,m} \rangle$, belonging to the
class of eigen-states specified by the quantum number $(n, m)$,
are generated by operating the $U (1)$ gauge transformation matrix
$U^{(\chi \leftarrow S)}$ on the symmetric eigen-state 
$\vert \Psi^{(S)}_{n, m} \rangle$,
where the same transformation matrix changes the gauge potential
configuration from $\bm{A}^{(S)}$ to $\bm{A}^{(\chi)}$, and similarly
for the other two classes of eigen-states. These three classes of
eigen-states belong to {\it different} or {\it inequivalent gauge class} in the 
sense that they are not connected through simple $U (1)$ gauge transformations. }
%\end{center}
\end{figure}

%%%%%%%%%%%%%%%%%%%%%%%%%%%%%%%%%%%%%%%%%%%%%%
%Probably, the confusion arises from the fact that the author of \cite{Govaerts2023} 
%fails to correctly recognize a delicate but important difference between the
%choice of the gauge potential (configuration) and the choice of gauge in the 
%quantum Landau problem. In the traditional formulation of the Landau problem,
%there is one-to-one correspondence between the choice of the
%gauge potential configuration and the choice of gauge, or more precisely,
%the choice of the corresponding Landau eigen-states. 
%On the other hand, in the gauge-potential-independent formulation,
%the choice of the gauge potential configuration does not necessarily mean 
%complete fixing of gauge.  
%In fact, from our analysis so far, one should clearly recognize 
%the fact that any specific gauge potential configuration 
%(including the symmetric gauge potential, the 1st Landau gauge potential,
%and the 2nd Landau gauge potential) appears {\it three times} in the different 
%types of Landau eigen-states, respectively having
%the rotational symmetry around the $z$-axis, the translational symmetry
%with respect to the $x$-axis, and the translational symmetry with
%respect to the $y$-axis.
%We identify these three type of
%eigen-states as belonging to different gauge classes. 
%On the other hand, the author of \cite{Govaerts2023} claims that, in the 
%Landau problem, there is only one gauge equivalence 
%class.\footnote{The gist is that
%this viewpoint might be valid in classical Landau problem, but it
%is not correct in quantum Landau problem, which we are concerned with.}

\vspace{2mm}
After the preparations above, we are ready to show that the remarkable claim (iii) 
made in the paper \cite{Govaerts2023} is not also justified.
An important tool here is the famous {\it gauge principle},
which dictates that the expectation values of gauge-invariant quantities must
be independent of the gauge choice.
As extensively discussed in \cite{WH2022}, what is crucial here is the comparison of 
the {\it quantum expectation values} of the
conserved quantities $\hat{p}^{cons}_x (\bm{A})$, $\hat{L}^{cons}_z (\bm{A})$ 
and those of the mechanical quantities $\hat{p}^{mech}_x (\bm{A})$, $\hat{L}^{mech}_z (\bm{A})$
between the two {\it different gauge classes} of eigen-state $\vert \Psi^{(\chi)}_{n, m} \rangle$
and $\vert \Psi^{(\chi^\prime)}_{n, k_x} \rangle$.
Before comparing the expectation values of the above four operators,
some technical preparation is mandatory. 
Since the 2nd type of eigen-states have a plane-wave-like normalization (with respect
to the $x$-direction) given as
\begin{equation}
 \int\!\!\!\int d x \,d y \, \Psi^{(\chi^\prime)^*}_{n, k^\prime_x} (x, y) \,
 \Psi^{(\chi^\prime)}_{n, k_x} (x, y) = \delta (k^\prime_x - k_x) ,
\end{equation}
their expectation values (or the {\it diagonal} matrix elements) of any operator
{\it diverge}. It is therefore necessary to introduce the corresponding normalizable
eigen-functions with the wave-packet-like nature \cite{WH2022}. 
They are obtained by the following
replacement of the plane-wave-like part of the eigen-functions
$\Psi^{(\chi^\prime)}_{n, k_x} (x, y)$
\begin{equation}
 \frac{1}{\sqrt{2 \,\pi}} \,e^{\,i \,k_x \,x} \ \rightarrow \  
 F_{k_x} (x) = \int_{- \,\infty}^\infty \frac{d k}{\sqrt{2 \,\pi}} \,\,
 g (k - k_x) \,e^{\, i \,k \,x} ,
\end{equation}
where $g(k)$ is an appropriate weight function of superposition, which
is supposed to have a peak at $k = 0$ and is normalized as
\begin{equation}
 \int_{- \,\infty}^\infty \,d k \,\,\vert g (k) \vert^2 \ = \ 1.
\end{equation}
Corresponding to the above replacement, the eigen-functions
above are replaced by
\begin{equation}
 \Psi^{(\chi^\prime)}_{n, k_x} (x, y) \, \rightarrow \, 
 \tilde{\Psi}^{(\chi^\prime)}_{n, k_x} (x, y) \, = \, F_{k_x} (x) \,Y_n (y),
\end{equation}
with
\begin{equation}
 Y_n (y) \ = \ N_n \,H_n \left( \frac{y - y_0}{l_B} \right) \,
 e^{\,- \,\frac{(y - y_0)^2}{2 \,l^2_B}} .
\end{equation}
It can be easily verified that they have the following normalization :  
\begin{equation}
 \int\!\!\!\int \, d x \, d y \,\,\,\tilde{\Psi}^{(\chi^\prime)^*}_{n, k_x} (x, y) \,
 \tilde{\Psi}^{(\chi^\prime)}_{n, k_x} (x, y) \ = \ 1.
\end{equation}

To proceed, we recall an important observation made in \cite{WH2022}.
As is widely known, the mechanical momentum operator
$\hat{p}^{mech}_x$ as well as the mechanical OAM operator
$\hat{L}^{mech}_z$ transform covariantly under a quantum mechanical
gauge transformation. Curiously, this is also the case
with the conserved momentum operator $\hat{p}^{cons}_x$ and
the conserved OAM operator $\hat{L}^{cons}_z$.
(This is easily convinced from the definition of $\hat{p}^{cons}_x$ given
in (\ref{def_px_cons}) and that of $\hat{L}^{cons}_z$ given in (\ref{def_Lz_cons}).)
An immediate consequences of this covariant gauge transformation
property of the mechanical and conserved operators are the 
following identities  :   
\begin{eqnarray}
 &\,& \langle \tilde{\Psi}^{(\chi^\prime)}_{n, k_x} \,
 \vert \,\hat{p}^{mech}_x (\bm{A}^{(\chi^\prime)}) \,
 \vert \tilde{\Psi}^{(\chi^\prime)}_{n, k_x} \rangle \ = \  
 \langle \tilde{\Psi}^{(\chi)}_{n, k_x} \,
 \vert \,\hat{p}^{mech}_x \,(\bm{A}^{(\chi)}) \,
 \vert \tilde{\Psi}^{(\chi)}_{n, k_x}  \rangle , \\
 &\,& \langle \tilde{\Psi}^{(\chi^\prime)}_{n, k_x} 
 \,\vert \,\hat{L}^{mech}_z \,(\bm{A}^{(\chi^\prime)} ) \,
 \vert \tilde{\Psi}^{(\chi^\prime)}_{n, k_x} \rangle \ = \  
 \langle \tilde{\Psi}^{(\chi)}_{n, k_x} \,\vert \,
 \hat{L}^{mech}_z \,(\bm{A}^{(\chi)} ) \,
 \vert \tilde{\Psi}^{(\chi)}_{n, k_x} \rangle , \\
 &\,& \langle \Psi^{(\chi^\prime)}_{n, m} \,\vert \,
 \hat{p}^{mech}_x \,(\bm{A}^{(\chi^\prime)} ) \,
 \vert \Psi^{(\chi^\prime)}_{n, m} \rangle \ = \  
 \langle \Psi^{(\chi)}_{n, m} \,\vert \,
 \hat{p}^{mech}_x \,(\bm{A}^{(\chi)} ) \,
 \vert \Psi^{(\chi)}_{n, m} \rangle , \\
 &\,& \langle \Psi^{(\chi^\prime)}_{n, m} \,\vert \,
 \hat{L}^{mech}_z \,(\bm{A}^{(\chi^\prime)} ) \,
 \vert \Psi^{(\chi^\prime)}_{n, m} \rangle \ = \  
 \langle \Psi^{(\chi)}_{n, m} \,\vert \,
 \hat{L}^{mech}_z \,(\bm{A}^{(\chi)} ) \,
 \vert \Psi^{(\chi)}_{n, m} \rangle , \hspace{10mm}
\end{eqnarray}
and
\begin{eqnarray}
 &\,& \langle \tilde{\Psi}^{(\chi^\prime)}_{n, k_x} \,\vert \,
 \hat{p}^{cons}_x \,(\bm{A}^{(\chi^\prime)} ) \,
 \vert \tilde{\Psi}^{(\chi^\prime)}_{n, k_x} \rangle \ = \  
 \langle \tilde{\Psi}^{(\chi)}_{n, k_x} \,\vert \,
 \hat{p}^{cons}_x \,(\bm{A}^{(\chi)} ) \,
 \vert \tilde{\Psi}^{(\chi)}_{n, k_x} \rangle , \\
 &\,& \langle \tilde{\Psi}^{(\chi^\prime)}_{n, k_x} \,\vert \,
 \hat{L}^{cons}_z \,(\bm{A}^{(\chi^\prime)} ) \,
 \vert \tilde{\Psi}^{(\chi^\prime)}_{n, k_x} \rangle \ = \  
 \langle \tilde{\Psi}^{(\chi)}_{n, k_x} \,\vert \,
 \hat{L}^{cons}_z \,(\bm{A}^{(\chi)} ) \,
 \vert \tilde{\Psi}^{(\chi)}_{n, k_x} \rangle , \\
 &\,& \langle \Psi^{(\chi^\prime)}_{n, m} \,\vert \,
 \hat{p}^{cons}_x \,(\bm{A}^{(\chi^\prime)} ) \,
 \vert \Psi^{(\chi^\prime)}_{n, m} \rangle \ = \  
 \langle \Psi^{(\chi)}_{n, m} \,\vert \,
 \hat{p}^{cons}_x \,(\bm{A}^{(\chi)} ) \,
 \vert \Psi^{(\chi)}_{n, m} \rangle , \\
 &\,& \langle \Psi^{(\chi^\prime)}_{n, m} \,\vert \,
 \hat{L}^{cons}_z \,(\bm{A}^{(\chi^\prime)} ) \,
 \vert \Psi^{(\chi^\prime)}_{n, m} \rangle \ = \  
 \langle \Psi^{(\chi)}_{n, m} \,\vert \,
 \hat{L}^{cons}_z \,(\bm{A}^{(\chi)} ) \,
 \vert \Psi^{(\chi)}_{n, m} \rangle . \hspace{10mm}
\end{eqnarray}
Here, $\chi^\prime$ and $\chi$ are arbitrary gauge function characterizing
the vector potential configuration in each
of the two types of eigen-states, which diagonalizes either of
$(\hat{H}, \,\hat{p}^{cons}_x)$ or $(\hat{H}, \,\hat{L}^{cons}_z)$.
The above equalities are naturally expected to hold, because both of
the mechanical quantities and the conserved quantities are
thought to be gauge-invariant quantities as long
as the expectation values {\it within the same gauge class} of eigen-states
are concerned.

\vspace{2mm}
What is highly nontrivial is the comparison of the expectation values of the conserved 
and mechanical operators between the eigen-states belonging to two {\it different gauge classes}. 
For the mechanical quantities, it was shown in \cite{WH2022} that
\begin{eqnarray}
 &\,& \langle \tilde{\Psi}^{(\chi^\prime)}_{n, k_x} \,\vert \,
 \hat{p}^{mech}_x (\bm{A}^{(\chi^\prime)})
 \,\vert \tilde{\Psi}^{(\chi^\prime)}_{n, k_x} \rangle \, = \,  
 \langle \Psi^{(\chi)}_{n, m} \,\vert \,
 \hat{p}^{mech}_x (\bm{A}^{(\chi)})
 \,\vert \Psi^{(\chi)}_{n, m} \rangle
 \, = \, 0 , \label{EV_pmech} \\
 &\,& \langle \tilde{\Psi}^{(\chi^\prime)}_{n, k_x} \,\vert \,
 \hat{L}^{mech}_z (\bm{A}^{(\chi^\prime)})
 \,\vert \tilde{\Psi}^{(\chi^\prime)}_{n, k_x} \rangle \, = \,  
 \langle \Psi^{(\chi)}_{n, m} \,\vert \,
 \hat{L}^{mech}_z (\bm{A}^{(\chi)})
 \,\vert \Psi^{(\chi)}_{n, m} \rangle
 =  (2 \,n + 1). \label{EV_Lmech} \hspace{10mm}
\end{eqnarray}
On the contrary, for the conserved quantities, it turned out that \cite{WH2022}
\begin{eqnarray}
 &\,& \langle \tilde{\Psi}^{(\chi^\prime)}_{n, k_x} \,\vert \,
 \hat{p}^{cons}_x (\bm{A}^{(\chi^\prime)})
 \,\vert \tilde{\Psi}^{(\chi^\prime)}_{n, k_x} \rangle \ \neq \  
 \langle \Psi^{(\chi)}_{n, m} \,\vert \,
 \hat{p}^{cons}_x (\bm{A}^{(\chi)})
 \,\vert \Psi^{(\chi)}_{n, m} \rangle , \label{EV_pcons} \\
 &\,& \langle \tilde{\Psi}^{(\chi^\prime)}_{n, k_x} \,\vert \,
 \hat{L}^{cons}_z (\bm{A}^{(\chi^\prime)})
 \,\vert \tilde{\Psi}^{(\chi^\prime)}_{n, k_x} \rangle \ \neq \  
 \langle \Psi^{(\chi)}_{n, m} \,\vert \,
 \hat{L}^{cons}_z (\bm{A}^{(\chi)})
 \,\vert \Psi^{(\chi)}_{n, m} \rangle .\label{EV_Lcons} \hspace{6mm}
\end{eqnarray}

Now, a {\it vital question} is whether one regards this remarkable difference 
between the expectation values of the mechanical and conserved quantities 
just as an {\it accident or not}. 
Concerning the inequalities (\ref{EV_pcons}) and (\ref{EV_Lcons}) for the 
expectation values of the {\it conserved quantities} between the two different 
classes of Landau eigenstate, someone might take the viewpoint that  
there is no reason that the expectation values should agree because 
they are the expectation values with respect to different basis states
characterized by the different quantum numbers $(n, m)$ and $(n, k_x)$.
This explanation may sound reasonable as long as the conserved quantities
are concerned. However, if wants to adopt such an interpretation, one is obliged 
to explain the reason why the expectation values of the 
{\it mechanical quantities} with respect to different bases perfectly 
agree as Eqs. (\ref{EV_pmech}) and (\ref{EV_Lmech}) show.
(We emphasize that this explanation must be made in terms of simple or 
universal terminology of physics.)

\vspace{1mm}
Without doubt, the difference pointed out above cannot be a mere accident. 
In fact, the observation above is perfectly consistent
with the common belief in the physics community that the
mechanical quantities are {\it genuinely} gauge-invariant physical quantities
in the sense that its expectation value is independent of the choice of gauge. 
Note that this interpretation is justified just because we assign the 
above-mentioned three families of eigen-states as belonging to three
{\it different gauge classes}. 
In this viewpoint, then, the disagreement between
the expectation values of the conserved quantities can be interpreted as 
showing the fact that they are {\it not} genuinely gauge-invariant quantities,
despite their {\it formal} gauge-covariant transformation property.
(The argument in this section is mainly devoted to clarifying the difference between
the mechanical quantities and the conserved quantities. In appendix A, we shall give
some supplementary remarks for demystifying the intricate relation between the 
canonical quantities and the conserved quantities.)

\vspace{1mm}
In this way, in sharp contrast to the claim in \cite{Govaerts2023}, 
we confirm once again that there is a physically unmissable difference between 
the mechanical quantities
and the conserved quantities in the Landau problem, despite the fact
that {\it both} transform {\it  covariantly} under a gauge transformation.
We conjecture that the ultimate physical origin of this difference can be 
understood from the following general consideration.
As is widely known, the Lagrangian of the quantum electrodynamics (QED)
is obtained from the free Dirac Lagrangian
\begin{equation}
 {\cal L}_D \ = \ \bar{\psi} (x) \,( i \,\gamma^\mu \,\partial_\mu - m ) \,\psi (x),
\end{equation}
based on the $U(1)$ gauge-invariance requirement, and it is given as
\begin{equation}
 {\cal L}_{QED}  =  \bar{\psi} (x) \,( i \,\gamma^\mu \,D_\mu -  m ) \,\psi (x)
 \, - \, \frac{1}{4} \,F^{\mu \nu} (x) \,F_{\mu \nu} (x),
\end{equation}
with $D_\mu \equiv \partial_\mu \, + \, i \,e \,A_\mu (x)$. 
To be more precise, what plays a critical role in this construction
is the celebrated {\it minimal principle}, which demands the replacement of the 
ordinary derivative in the Dirac Lagrangian by the covariant derivative as
\begin{equation}
 \partial_\mu \ \rightarrow \ \partial_\mu \ + \ i \,e \,A_\mu (x) 
 \ \equiv \ D_\mu.
\end{equation}
Although this minimal prescription ensures the gauge invariance of the QED 
Lagrangian, the converse is not necessarily true. In fact,
the requirement of the gauge-invariance alone never prevents from
adding to the QED Lagrangian the term like
\begin{equation}
 \bar{\psi} (x) \,\sigma^{\mu \nu} \,\psi (x) \,F_{\mu \nu} (x),
\end{equation}
with $\sigma^{\mu \nu} \equiv \frac{i}{2} [\gamma^\mu, \gamma^\nu ]$. 
This term is certainly $U(1)$ gauge invariant, but we know that an
addition of such a term to the basic QED Lagrangian is not supported by nature.

\vspace{1mm}
At this point, we think it useful to recall the fact that both of the mechanical momentum
and the mechanical OAM can be related to the {\it gauge-invariant electric
current} of the electron, which is obtained from the {\it minimal prescription}.
On the other hand, the conserved momentum and the conserved OAM
are not such quantities that are directly related to the {\it minimal current}
of the electron.
Probably, the {\it genuinely} gauge-invariant (or  {\it observable}) nature of 
the mechanical quantities as compared
with the conserved quantities can be understood by this difference.   
In the following two sections, we try to show the validity of this
conjecture with the help of concrete examples of illustrative nature.
Those demonstrations are also expected to elucidate the physical reason why
the quantum numbers $k_x$ and $m$ are not direct experimental observables, 
even though they are important quantum numbers characterizing 
the Landau eigen-states in both of the traditional formulation and
the gauge-potential-independent formulation.

%%%%%%%%%%%%%%%%%%%%%%%%%%%%%%%%%%%%%%%%%
%%%%%%%%%%%%%%%%%%%%%%%%%%%%%%%%%%%%%%%%%
\section{Nonobservable nature of the quantum number $k_x$ characterizing
the 1st Landau gauge eigen-states}
\label{sec4}

The observations made in sect.\ref{sec2} revealed the fact that the
quantum number $k_x$ appearing in the eigen-functions in the 1st
Landau gauge has little to do with the motion in the $x$-direction,
but instead it is related to the $y$-coordinate $y_0$ of the guiding center 
of the cyclotron motion related through $y_0 = l^2_B \,k_x$. 
On the other hand,
the $x$-coordinate of the guiding center is totally uncertain and
it is distributed uniformly along the line $y = y_0$.
This strongly indicates that the quantum number $k_x$ is not such a
quantity that would be related to direct observables. 
In the following, we try to demonstrate that this is indeed
the case, through the consideration of the familiar Hall effect as a 
concrete and instructive example \cite{Murayama-Berkley, Jaffe-MIT}.

The Hall effect is observed when the electric field $E$ is additionally applied
to the perpendicular direction (say the $y$-axis direction) to the $z$-axis, which
is the direction of the uniform magnetic field. The relevant Hamiltonian
is given as
\begin{eqnarray}
 H^\prime &=& \frac{1}{2 \,m_e} \,( \Pi^2_x \,+ \,\Pi^2_y ) \, - \, 
 e \,E \,y \nonumber \\
 &=& \frac{1}{2 \,m_e} \left\{ ( p^{can}_x + e \,A_x )^2 + 
 ( p^{can}_y + e \,A_y )^2 \right\}  \ - \ e \,E \,y .
\end{eqnarray}
Here, $p^{can}_x$ and $p^{can}_y$ stand for the {\it canonical momentum}
operators. This problem is convenient to treat in the 1st Landau gauge
$\bm{A}^{(L_1)} = B \,( - \,y, 0)$. With this gauge choice, $H^\prime$ reduces to
\begin{eqnarray}
 H^\prime &=& \frac{1}{2 \,m_e} \left\{ ( p^{can}_x \,- \,e \,B \,y )^2 
 \,+ \,( p^{can}_y )^2 \right\}  \ -\ e \,E \,y.
\end{eqnarray}
Since the Hamiltonian $H^\prime$ does not explicitly contain the coordinate $x$,
it has eigen-functions of the following form : 
\begin{equation}
 \Psi^{\prime (L_1)} (x, y) \ \propto \ e^{\,i \,k_x \,x} \,\,Y^\prime (y),
\end{equation}
so that the canonical momentum operator $p^{can}_x$ can effectively be replaced
by a $c$-number $k_x$. Then, $H^\prime$ can effectively be transformed as
\begin{eqnarray}
 H^\prime &\rightarrow&
 \frac{1}{2 \,m_e} \left\{ (k_x \,- \,e \,B \,y )^2 \,+ \,
 (p^{can}_y )^2  \right\} \ - \ e \,E \,y \nonumber \\
 \!&=&\! \frac{1}{2 \,m_e} \left\{ ( p^{can}_y )^2 \,+ \,
 \left( k_x \,- \,e \,B \,y \,- \,m_e \,\frac{E}{B} \right)^2 \right\} 
  \, + \, \mbox{constant} \nonumber \\
 \!&=&\! \frac{1}{2 \,m_e} \Biggl\{ ( p^{can}_y )^2 \Biggr.  + (e \,B )^2 \left. \left[\,
 y - \frac{1}{e \,B} \,\left( k_x  +  m_e \,\frac{E}{B} \right) \right]^2
 \right\}   +  \mbox{constant} .  \hspace{8mm}
\end{eqnarray}
Now, the eigen-functions of the above Hamiltonian can easily be obtained 
as
\begin{equation}
 \Psi^{\prime (L_1)}_{n, k_x} (x, y) \ = \ \frac{e^{\,i \,k_x \,x}}{\sqrt{2 \,\pi}} \,\,
 Y^\prime_n (y) .
\end{equation}
where
\begin{equation}
 Y^\prime_n (y) \ = \ N_n \,H_n \left( \frac{y - y^\prime_0}{l_B} \right) \,
 e^{\,- \,\frac{(y - y^\prime_0)^2}{2 \,l^2_B}} ,
\end{equation}
with
\begin{equation}
 y^\prime_0 \ = \ \frac{k_x}{e \,B} \ + \ \frac{m_e E}{e \,B^2} .
\end{equation}

\vspace{1mm}
Since these eigen-functions have a plane-wave-like normalization
(with respect to the $x$-axis), it is convenient to introduce 
the corresponding normalizable eigen-functions with wave-packet-like
nature, as was done in the previous section.
This is achieved by the following replacement : 
\begin{eqnarray}
 \Psi^{\prime (L_1)}_{n, k_x} (x, y) \rightarrow  
 \tilde{\Psi}^{\prime (L_1)}_{n, k_x} (x, y)  = F_{k_x} (x) \,Y^\prime_n (y),
 \hspace{4mm}
\end{eqnarray}
where the function $F_{k_x} (x)$ is defined in the previous section.

Now we are ready to continue our discussion.
As is widely known, the gauge-invariant probability current density of
the electron is represented by the velocity operator or the mechanical
momentum operator as
\begin{eqnarray}
 \bm{j} \, = \, e \,\bm{v} &=& \frac{e}{m_e} \,\Psi^* \,\bm{p}^{mech} \,\Psi
 \ = \ \frac{e}{m_e} \,\Psi^* \,\left( \bm{p}^{can} \,+ \,e \,\bm{A} \right) \,\Psi ,  
 \hspace{8mm}
\end{eqnarray}
where $\Psi$ is the electron wave function in some gauge.
Since we are working in the 1st Landau gauge, the corresponding
expectation values of $j_x$ and $j_y$ become
\begin{eqnarray}
 \langle j_x \rangle &=& \frac{e}{m_e} \int\!\!\!\int dx \,dy \,\,
 \tilde{\Psi}^{\prime (L_1)^*}_{n, k_x} (x, y) \left( p^{can}_x - e \,B \,y
 \right) \tilde{\Psi}^{\prime (L_1)}_{n, k_x} (x, y), \hspace{6mm}
 \label{EV_jx} \\
 \langle j_y \rangle &=& \frac{e}{m_e} \,\int\!\!\!\int dx \,dy \,\,
 \tilde{\Psi}^{\prime (L_1)^*}_{n, k_x} (x, y) \,\,p^{can}_y \,\,
 \tilde{\Psi}^{\prime (L_1)}_{n, k_x} (x, y) .
\end{eqnarray}
First, it is easy to verify the relation,
\begin{eqnarray}
 \langle j_y \rangle &=& \frac{e}{m_e} \int\!\!\!\int dx \,dy \,
 \tilde{\Psi}^{\prime (L_1)^*}_{n, k_x} (x, y) \,p^{can}_y \,
 \tilde{\Psi}^{\prime (L_1)}_{n, k_x} (x, y) \nonumber \\
 &=& \frac{e}{m_e} \int d y \,\, Y^\prime_n (y) \,p^{can}_y \,Y^\prime_n (y)
 \ = \ 0.
\end{eqnarray} 
Next, we obtain
\begin{equation}
 \langle j_x \rangle \ = \ \langle j^{can}_x \rangle \ + \ 
 \langle j^{gauge}_x \rangle ,
\end{equation}
with
\begin{eqnarray}
 \langle j^{can}_x \rangle &=& \frac{e}{m_e} \,\int\!\!\!\int \,dx \,dy \,\,
 \tilde{\Psi}^{\prime (L_1)^*}_{n, k_x} (x, y) \,\,p^{can}_x \,\,
 \tilde{\Psi}^{\prime (L_1)}_{n, k_x} (x, y)  , \\
 \langle j^{gauge}_x \rangle &=& \frac{e}{m_e} \,\int\!\!\!\int \,dx \,dy \,\,
 \tilde{\Psi}^{\prime (L_1)^*}_{n, k_x} (x, y) \,\,( - \,e \,B \, y ) \,\,
 \tilde{\Psi}^{\prime (L_1)}_{n, k_x} (x, y)  . \hspace{6mm}
\end{eqnarray}
Using the relation
\begin{equation}
 \int \,dx \,\,F^*_{k_x} (x) \,p^{can}_x \,F_{k_x} (x) \ = \ k_x,
\end{equation}
together with the relation
\begin{equation}
 \int \,dy \,\,\left[ Y^\prime_n (y) \right]^2 \ = \ 1,
\end{equation}
we find that
\begin{eqnarray}
 \langle p^{can}_x \rangle \!&=&\! 
 \int\!\!\!\int dx \,dy \,\tilde{\Psi}^{\prime (L_1)^*}_{n, k_x} (x, y) \,
 p^{can}_x \,
 \tilde{\Psi}^{\prime (L_1)}_{n, k_x} (x, y) \ = \ k_x .
\end{eqnarray}
The expectation value of $y$, which appears
in Eq.(\ref{EV_jx}), can be calculated as follows.
\begin{eqnarray}
 \langle y \rangle &=& \int\!\!\!\int dx \,dy \,\,
 \tilde{\Psi}^{\prime (L_1)^*}_{n, k_x} (x, y) \,\,y \,\,
 \tilde{\Psi}^{\prime (L_1)}_{n, k_x} (x, y) \ = \ 
 \int dy \,\,Y^\prime_n (y) \,y \,Y^\prime_n (y)  \hspace{5mm} \nonumber \\
 &=& \int dy \,\,Y^\prime_n (y) \,( y \,- \,y^\prime_0 \,+ \,y^\prime_0 ) \,
 Y^\prime_n (y) \ = \  \ y^\prime_0  \ = \ 
 \frac{k_x}{e \,B} \ + \ \frac{m_e E}{e \,B^2} .
\end{eqnarray}
In this way, we eventually get
\begin{eqnarray}
 \langle j^{can}_x \rangle &=& \frac{e}{m_e} \,k_x , \\
 \langle j^{gauge}_x \rangle &=& - \,\frac{e}{m_e} \,k_x
 \ - \ e \,\,\frac{E}{B} , \hspace{5mm}
\end{eqnarray}
which in turn gives
\begin{eqnarray}
 \langle j_x \rangle &=& 
 \langle j^{can}_x \rangle \ + \ \langle j^{gauge}_x \rangle
 \nonumber \\
 &=&  
 \frac{e}{m_e} \,k_x \ + \ 
 \left( - \,\frac{e}{m_e} \,k_x \ - \,e \,\,\frac{E}{B} \right) \ = \ 
  - \,e \,\,\frac{E}{B} .
\end{eqnarray}
Accordingly, the expectation value of the velocity operator
$v_x$ becomes
\begin{equation}
 \langle v_x \rangle \ = \ \frac{1}{e} \,\langle j_x \rangle
 \ = \ - \,\frac{E}{B} .
\end{equation}
This is nothing but the familiar {\it drift velocity} of the electron
in the Hall effect, which obviously corresponds to a direct
observable \cite{Murayama-Berkley, Jaffe-MIT}.
All these affairs are nothing new and they are textbook materials
of the non-relativistic quantum mechanics.  
To our knowledge, however, this material has seldom been
discussed from the question about the observability or non-observability
of the quantum number $k_x$ characterizing the Landau eigen-states.
One can confirm that the quantum number $k_x$, which is the
eigen-values of the canonical momentum operator $p^{can}_x$,
certainly appears in the expectation values of the canonical part and also
in the gauge-potential-dependent part of the net probability current
operator $j_x$. However, in the sum of the canonical and
gauge part, they cancel out exactly.
This means that, the canonical and gauge parts of the net current 
(i.e. the mechanical current) cannot be separately observed,
which is equivalent to say that the quantum number $k_x$ does
not correspond to a direct experimental observable.\footnote{One
must recognize the fact that nonobservability of the quantum number
$k_x$ means not only nonobservability of the canonical momentum
but also nonobservability of conserved momentum, since $k_x$ is
also the eigenvalue of the latter operator.}
It seems to us that this is not only consistent with the gauge principle
that demands non-observability of the gauge-variant canonical
momentum but also the nature of the quantity $k_x$ as clarified in
sect.2.
In fact, we have shown there that the quantum number $k_x$ has little to
do with the motion in the $x$-direction, but instead it rather
represents the $y$-coordinate of the guiding center, which is uniformly
distributed (in a quantum mechanical probability sense) along the 
line $y = y_0 \equiv l^2_B \,k_x$. It may be better to emphasize that 
most of these observations are gauge-dependent one, which holds only 
within choice of the 1st Landau gauge. However, it is a universally
correct fact that the net mechanical momentum is a gauge-invariant 
quantity and that it correspond to direct observable.

\section{Nonobservable nature of the quantum number $m$ 
characterizing the symmetric gauge Landau eigen-states}
\label{sec5}

It is well-known that there is an intimate connection between the Landau
Hamiltonian and the Hamiltonian of the 2-dimensional 
Harmonic oscillator \cite{JL1949}.
In particular, the magnetic quantum number $m$ appears in both of the
eigen-states of the 2-dimensional Harmonic oscillator in the spherical basis
and the eigen-states of the Landau Hamiltonian in the symmetric gauge.
Here, we confront the observability or non-observability question of the 
magnetic quantum number $m$ in both systems by paying a special attention
to the similarities and dissimilarities of these two quantum mechanical 
systems \cite{Tiwari2023}. 
First, the Hamiltonian of the
isotropic 2-dimensional Harmonic oscillator is given by
\begin{equation}
 H_{osc} \ = \ \frac{1}{2 \,m} \,( p^2_x \,+ \,p^2_y ) \ + \ \frac{1}{2} \,m \,\omega^2 \,
 ( x^2 + y^2) .
\end{equation}
(In this section, the canonical momentum operators are simply denoted
as $p_x$ and $p_y$ to avoid unnecessary notational complexity.) 
The problem of the 2-dimensional Harmonic oscillator can be solved in
either of the Cartesian coordinate system or the spherical coordinate system.
Let us start the discussion with the eigen-states of the 2-dimensional
Harmonic oscillator in the Cartesian basis. 
As is well-known, these eigen-states are characterized by two non-negative
integers $n_x$ and $n_y$ and represented as 
(see, for example, \cite{Cohen-Tannoudji2020})
\begin{equation}
 \vert n_x, n_y \rangle \ = \ \frac{1}{\sqrt{n_x ! \,n_y !}} \,\,
 (a^\dagger_x)^{n_x} \,(a^\dagger_y)^{n_y} \,\,\vert 0, 0 \rangle .
\end{equation}
Here, $a^\dagger_x$ and $a^\dagger_y$ are the creation operators
of the Harmonic oscillator quanta with respect to the $x$- and
$y$-directions, while the state $\vert n_x = 0, n_y = 0 \rangle$ stands for
the vacuum of the Harmonic oscillator quanta. 
The states $\vert n_x, n_y \rangle$ satisfy the following eigen-value equation : 
\begin{equation}
 H_{osc} \,\vert \,n_x, n_y \rangle \ = \ E \,\vert \,n_x, n_y \rangle ,
\end{equation}
with the eigen-energies
\begin{equation}
 E \ = \ \left( n_x \,+ \, n_y \, + \, 1 \right) \,\omega . \hspace{4mm}
\end{equation}
Note that, by introducing the quantum
number $n$ by $n \equiv n_x + n_y$, the above eigen-energies can also be
expressed as $E = ( n + 1) \,\omega$. It is easy to
verify that the state with the eigen-energy $( n + 1) \,\omega$ has 
$( n + 1 )$-fold degeneracy.

The same problem can be solved also in the spherical basis.
The eigen-states in the spherical basis are characterized by
two integers $n$ and $m$, and represented as
$\vert n, m \rangle$. They are the simultaneous eigen-states of
the Hamiltonian $H_{osc}$ and the canonical OAM operator 
$L_z = x \,p_y - y \,p_x$, which satisfy the eigen equations : 
\begin{eqnarray}
 H_{osc} \,\vert n, m \rangle &=& ( n + 1) \,\omega \,\vert n, m \rangle, 
 \hspace{4mm} \\
 L_z \,\vert n, m \rangle &=& \ m \,\vert n, m \rangle .
\end{eqnarray}
As shown in Appendix B, for a fixed value of $n$, the magnetic quantum
number $m$ takes the following $(n + 1)$ values,
\begin{equation}
 m \ = \ n, \, n-2 \, \cdots , \, - \,(n-2), - \,n.
\end{equation}
Namely, one confirms that, also in the spherical-basis treatment,
the state with the eigen-energy $(n + 1) \,\omega$ has
$(n + 1)$-fold degeneracy. 

\vspace{2mm}
%\noindent
Now let us ask the following question.  
What happens if we add the uniform magnetic field along the
$z$-direction to the system of 2-dimensional Harmonic oscillator ? 
The perturbed Hamiltonian would be given in the form : 
\begin{equation}
 H^\prime_{osc} \ = \ H_{osc} \ + \ \Delta H, \ \ \mbox{with} \ \ 
 \Delta H \ = \ \lambda \,L_z,
\end{equation}
where
\begin{equation}
 \lambda \ = \ \mu_B \ \ \mbox{with} \ \ 
 \mu_B \ = \ \frac{q}{2 \,m} .
\end{equation}
In this problem, we assume that the particle moving in the 
Harmonic oscillator potential has a charge $q$.
The answer is very simple, if we consider the problem in the spherical
basis. Since the $\vert n, m \rangle$-basis is the eigen-states of $L_z$
with the eigen-value $m$, we immediately find that
\begin{equation}
 H^\prime_{osc} \,\vert n, m \rangle \ = \ E^\prime \,\vert n, m \rangle ,
\end{equation}
where
\begin{equation}
 E^\prime \ = \ ( n \,+ \,1 ) \,\omega \ + \ \lambda \,m .
\end{equation}
This means that, due to the uniform magnetic field applied along the
$z$-direction, the degeneracy about the quantum number $m$ is
lifted (Zeeman splitting). It clearly shows that the quantum
number $m$ characterizing the eigen-states of the 2-dimensional 
Harmonic oscillator corresponds to a {\it direct observable}. 

\vspace{1mm}
How can we answer the same question in the
treatment with the Cartesian basis $\vert n_x, n_y \rangle$?
At first, it seems to be a somewhat nontrivial question, because the 
magnetic quantum number $m$ never  appears in the Cartesian 
basis $\vert n_x, n_y \rangle$.
If one naively applies the non-degenerate perturbation theory
in the 1st order, the perturbed energy is given by
\begin{eqnarray}
 \Delta E^{(1)} &=&\langle n_x, n_y \,\vert \,
 \Delta H \,\vert n_x, n_y \rangle \nonumber \\
 &=& \lambda \,\langle n_x, n_y \,\vert \,
 i \,( a_x \,a^\dagger_y \,- \,
 a^\dagger_x \,a_y ) \,\vert n_x, n_y \rangle 
 \ = \ 0 . 
\end{eqnarray}
Therefore, it looks as if no Zeeman splitting occurs. 
It is of course wrong. Because there are
degeneracies about the energy eigen-states, we must apply the
degenerate perturbation theory. (In our present problem, the
following procedure is equivalent to diagonalizing anew 
the total Hamiltonian $H^\prime_{osc}$. We nevertheless think that the 
consideration below is instructive to understand the meaning and
the answer of our proposed question above.)

%\vspace{2mm}
Let us look into several eigen-states with smaller value of $n$ 
in order. First, there is no degeneracy in the $n = 0$ state.
Next, for $n = 1$, the two states $\vert n_x = 1, n_y = 0 \rangle$
and $\vert n_x = 0, n_y = 1 \rangle$ have the same energy.
Following the strategy of the degenerate perturbation theory, we
calculate the matrix elements of $\Delta H = \lambda \,L_z$
within the space of these two bases, and write down the secular equation. 
The matrix elements are given as
\begin{eqnarray}
 \langle n_x = 1, n_y = 0 \,\vert \,L_z \,\vert n_x = 1, n_y = 0 \rangle
 &=& 0, \\
 \langle n_x = 1, n_y = 0 \,\vert \,L_z \,\vert n_x = 0, n_y = 1 \rangle
 &=& - \,i, \hspace{8mm} \\ 
 \langle n_x = 0, n_y = 1 \,\vert \,L_z \,\vert n_x = 1, n_y = 0 \rangle
 &=& + \,i, \\
 \langle n_x = 0, n_y = 1 \,\vert \,L_z \,\vert n_x = 0, n_y = 1 \rangle
 &=& 0,
\end{eqnarray}
so that the corresponding secular equation becomes
\begin{equation}
 \left\vert \begin{array}{cc}
 \ 0 \ - \ E^{(1)} \ & \ - \,i \,\lambda \ \\
 \ i \,\lambda \ & \ 0 \ - \ E^{(1)} \\
 \end{array} \right\vert \ = \ 0.
\end{equation}
Solving the secular equation, one finds that the perturbed eigen-energies are
given by
\begin{equation}
 E^{(1)} \ = \ \pm \,\lambda ,
\end{equation}
while the corresponding eigen-vectors are given by
\begin{eqnarray}
 &\vert \psi^{(1)}_{n=1, \alpha} \rangle& \ = \ 
 \frac{1}{\sqrt{2}} \, 
 \left\{ \vert n_x = 1, n_y = 0 \rangle + i \,
 \vert n_x = 0, n_y = 1 \rangle \right\} , \hspace{8mm} \\
 &\vert \psi^{(1)}_{n=1, \beta} \rangle&\ = \ 
 \frac{1}{\sqrt{2}} \,
 \left\{ \vert n_x = 1, n_y = 0 \rangle \ - \ i \,
 \vert n_x = 0, n_y = 1 \rangle \right\}. \hspace{8mm}
\end{eqnarray}
Using the general formula relating the $\vert n_x, n_y \rangle$-basis
and the $\vert n, m \rangle$-basis derived in Appendix, one can
easily confirm that the two eigen-vectors obtained above are nothing
but the states $\vert n=1, m=1 \rangle$ and $\vert n=1, m = - \,1 \rangle$
in the spherical basis, respectively. 

\vspace{3mm}
Next, for $n=2$, the three states $\vert n_x = 2, n_y = 0 \rangle$,
$\vert n_x = 1, n_y = 1 \rangle$ and $\vert n_x = 0, n_y = 2 \rangle$
are energetically degenerate. By diagonalizing $\Delta H$ in these
bases, the perturbed eigen-energies are obtained as
\begin{equation}
 E^{(1)} \ = \ 2 \,\lambda, \,\, 0, \,\, - 2 \,\lambda ,
\end{equation}
and the corresponding eigen-vectors become
\begin{eqnarray}
 \vert n = 2, m = 2 \rangle \ &=& \ \frac{1}{2} \,\Bigl\{
 \vert n_x = 2, n_y = 0 \rangle 
 - \vert n_x = 0, n_y = 2 \rangle \Bigr.  \nonumber \\
 &\,& \hspace{35mm} \Bigl. + \ \sqrt{2} \,i \,
 \vert n_x = 1, n_y = 1 \rangle \Bigr\} ,  \hspace{10mm} \\
 \vert n = 2, m = 0 \rangle \ &=& \ \frac{1}{2} \,\Bigl\{
 \vert n_x = 2, n_y = 0 \rangle + 
 \vert n_x = 0, n_y = 2 \rangle \Bigr\} , \\
 \vert n = 2, m = - \,2 \rangle \ &=& \ \frac{1}{2} \,\Bigl\{
 \vert n_x = 2, n_y = 0 \rangle 
 - \vert n_x = 0, n_y = 2 \rangle \Bigr.  \nonumber \\
 &\,& \hspace{35mm} \Bigl. - \ \sqrt{2} \,i \,
 \vert n_x = 1, n_y = 1 \rangle \Bigr\} ,  \hspace{10mm}
\end{eqnarray}
In this way, we confirm that, if the uniform magnetic field
is added to the Hamiltonian of the 2-dimensional Harmonic
oscillator, the Zeeman splitting occurs in both of 
the spherical-basis treatment and the Cartesian-basis treatment, and the
magnetic quantum number $m$ in principle correspond to an observable.
This is nothing surprising, because the physics should not
change depending on how one chooses the coordinate basis
to solve the problem.

\vspace{3mm}
As we shall see below, however, the situation is remarkably
different in the Landau problem.
Under the assumption of the uniform magnetic field directed to
the $z$-direction, the Landau problem is essentially a 2-dimensional
problem. However, a critical difference from the 2-dimensional
Harmonic oscillator system is that the Landau Hamiltonian {\it depends}
on the vector potential $\bm{A}$ as
\begin{equation}
 H_{Landau} \ = \ \frac{1}{2 \,m_e} \,\left(\, \bm{p} \ + \ e \,\bm{A} \,\right)^2 .
\end{equation}
The well-known problem is that the vector potential, which reproduces
the uniform magnetic field, is not unique (gauge arbitrariness).
The three popular gauge choices are the 1st Landau gauge, the 2nd Landau gauge,
and the symmetric gauge. For example, with the choice of the 1st
Landau gauge potential $\bm{A}^{(L_1)} = B \,( - \,y, 0)$, the Landau Hamiltonian
reduces to the form : 
\begin{equation}
 H_{Landau} = \frac{1}{2 \,m_e} \left\{ (p_x \,- \,e \,B \,y )^2 \, + \, p^2_y
 \right\} .
\end{equation}
On the other hand, with the choice of the symmetric gauge potential
$\bm{A}^{(S)} = \frac{1}{2} \,B \,( \,- \,y, \,x )$,
the Landau Hamiltonian reduces to
\begin{equation}
 H_{Landau} \ = \ H_{osc} \ + \ \omega_L \,L_z, 
\end{equation}
where $H_{osc}$ is the Hamiltonian of the 2-dimensional Harmonic
oscillator given by
\begin{equation}
 H_{osc} \ = \ \frac{1}{2 \,m_e} \,( p^2_x  +  p^2_y )  \ + \   
 \frac{1}{2} \,m_e \,\omega^2_L \,( x^2 + y^2) .
\end{equation}
In the above equation, $L_z \equiv x \,p_x \,- \,y \,p_x$ is 
the $z$-component of the canonical OAM operator,
whereas $\omega_L = \frac{e \,B}{2 \,m_e}$ is the familiar Larmor frequency.
What is important to recognize here is the fact that the Landau Hamiltonian
takes totally different forms depending on the choice of gauge potential.
Different from the problem of the simple 2-dimensional Harmonic oscillator,
this is {\it not} a mere difference of the choices of the Cartesian coordinate
system and the spherical coordinate system.
With the choice of the 1st Landau gauge, the Landau Hamiltonian shows
{\it translational symmetry} with respect to the $x$-axis.
On the other hand, with the choice of the symmetric gauge, the
Landau Hamiltonian shows {\it rotational symmetry} around the
coordinate origin. 
This should be contrasted with the fact that the Hamiltonian of the 
2-dimensional Harmonic oscillator has {\it rotational symmetry} 
around the coordinate origin {\it regardless of} the Cartesian coordinate 
treatment and the spherical coordinate treatment.

Although it is rarely mentioned, the difference between the two physical system
noted above is not unrelated to a very special nature of the 
Landau problem, in which the magnetic field is uniformly spreading over the whole
2-dimensional plane, so that the choice of the {\it coordinate origin} is totally
{\it arbitrary}. This should be contrasted with the problem of the 
2-dimensional Harmonic oscillator. 
In the latter problem, the choice of the coordinate origin
is practically unique. It is just the center of the Harmonic oscillator
potential. Note that this is usually the case also in other physical problems 
which we normally encounter.
For example, in the description of the hydrogen atom, the most natural 
choice of the coordinate origin is the position of the proton, which is much
heavier than the electron and is thought to be at rest in a good approximation. 
Though it is not usually mentioned clearly or explicitly, we claim that 
this very fact must be a crucial reason why the magnetic quantum 
numbers $m$ of the hydrogen eigen-states have good physical meaning as
degrees of intensity of rotational motion around a {\it definite orbit center}
and can be observed by means of the Zeeman splitting or others. 

Now we argue that this is not the case with the magnetic quantum number 
$m$, which appears in the Landau eigen-states in the symmetric gauge.
Certainly, the eigen-states of the Landau Hamiltonian in the
symmetric gauge is characterized by two quantum numbers, i.e. the
nonnegative Landau quantum number $n$ and the magnetic quantum
number $m$. 
Suppose that we attempt to lift the energy degeneracy in the
quantum number $m$ by imposing an additional magnetic field given as
\begin{equation}
 \Delta \bm{B} \ = \ \Delta B \,\bm{e}_z .
\end{equation}
The only changes caused by this additional magnetic field are the following : 
\begin{eqnarray}
 B &\rightarrow& B^\prime  \ =  \ B \, + \, \Delta B,  \\  
 \omega_L  &\rightarrow&  
 \omega^\prime_L  \ = \  \frac{e \,B^\prime}{2 \,m_e}, 
\end{eqnarray}
which means that the degeneracy in the quantum number $m$ is never lifted and
just remains. One therefore realizes that there is a crucial difference between
the physics of degeneracy in the magnetic quantum number $m$ 
in the Landau problem and that in the 2-dimensional Harmonic oscillator.
As mentioned above, the ultimate reason of this difference is thought to be 
connected with the complete arbitrariness in the choice of the 
{\it coordinate origin} in the Landau problem. 
In fact, this is reflected in the uncertainly of  the position
of the orbit center with respect to the coordinate origin hidden in the
Landau eigen-states even after we choose to work in the symmetric gauge.

\section{Implication on the gauge-invariant decomposition problem
of the nucleon spin}
\label{sec6}

Although it might appear that there is little connection between the discussions
in the present paper and the unsettled debates on the gauge-invariant decomposition
problem of the nucleon spin, there actually exists a deep connection between them.
The present section is devoted to briefly discussing this issue. 
The nucleon spin decomposition problem is how to gauge-invariantly
decompose the total spin of the nucleon, which is naturally one-half, to the contributions 
of the intrinsic spin parts and the orbital angular momentum parts of
quarks and gluons that make up the nucleon. 
It is widely recognized that there are two different types of decomposition :
one is the canonical-type decomposition characterized by the canonical OAM of 
quarks \cite{JM1990} and the other is the mechanical-type decomposition
characterized by the mechanical (or kinetic)  OAM of quarks inside the 
nucleon \cite{Ji1997, Waka2010}. 
(For earlier reviews, see \cite{LL2014, Waka2014}, for example.)
Originally, the canonical OAM of quarks appearing in the famous 
Jaffe-Manohar decomposition of the nucleon spin \cite{JM1990} was believed to 
be a  {\it gauge-variant} quantity.
However, after Chen et al.'s paper appeared \cite{Chen2008}, several authors 
proposed the concept of gauge-covariant (g.c.) extension of the canonical 
OAM \cite{Hatta2011, Lorce2013, LL2014}, and the belief that 
this {\it extended} canonical OAM can be regarded as a gauge-invariant quantity 
became popular. 
An apparent problem of such an idea is that the way of gauge-covariant extension
is far from unique and there are plural possibilities of 
extension \footnote{Generally speaking, the gauge symmetry is a redundancy, 
the extra gauge degrees of freedom of which should be 
eliminated to get physical predictions \cite{Schwartz2014, Zee2013}.
In that sense, it appears to us that the operation of 
gauge-invariant extension mistakes the means for the end at least from
the logical point of view.}
. 
Some popular examples are the g.c. extension based on the light-cone gauge, 
the g.c. extension based on the temporal gauge, the g.c. extension based on
the spatial axial gauge, and the g.c. extension based on the Coulomb gauge, etc. 
This implies that the way of g.c. extension depends on the choice of
{\it basis gauge}, and it in turn implies that g.c. extensions based on different 
basis gauges belong to {\it different gauge classes}.
We also remind of the fact that the above-mentioned gauge-invariant 
(or gauge-covariant) extension is realized by the introduction of the
non-local Wilson line. Although it may be formally gauge-invariant,
it is known that the non-local realization of the gauge-invariance should be
taken with caution \cite{Schwartz2014}.\footnote{In this textbook, it is stated
that, although  gauge invariance is merely a redundancy of description,
it makes it a lot easier to study field theory.  At first sight, it appear to
indicate mathematically useful but physically vacuous nature of the gauge 
invariance concept. However, we should not forget irreplaceable role of the 
gauge principle, which mediate between gauge-invariance concept
and observability. } 
For example, formally
gauge-invariant definition of the gluon spin operator also requires
non-local gauge link or Wilson line. On the other hand, however, it is a long 
known fact that there is no definition of the gluon spin, which satisfies
{\it locality} and the {\it gauge-invariance} at the same time.
This leaves the question whether the gluon spin
defined with the use of the non-local Wilson line is truly a gauge-invariant
quantity or not.

The significance of the present paper is to elucidate these intricate issues
related to the gauge theory by making full use of the analytically solvable 
nature of the Landau problem.
(See also the discussion in \cite{WKZ2018}, which is 
developed based on the gauge-invariant but path-dependent formulation 
{\it a la} DeWitt \cite{DeWitt1962} of the Landau problem.)
In fact, we pointed out that the conserved momentum
$\hat{p}^{cons}_x$ in our paper can be regarded as a g.c. extension of
the canonical momentum $\hat{p}^{can}_x$ based on the 1st Landau gauge,
while the conserved OAM $\hat{L}^{cons}_z$ in our paper can be thought of
as a g.c. extension of the canonical OAM $\hat{L}^{can}_z$ based on the
symmetric gauge.  We also elucidated the fact that, despite the formal
gauge-covariant transformation properties, neither $\hat{p}^{cons}_x$
nor $\hat{L}^{cons}_z$ can be regarded as genuinely gauge-invariant
physical quantities, which is critically different from the mechanical 
quantities like $\hat{p}^{mech}_x$ and $\hat{L}^{mech}_z$.

It may be interesting to compare these two types of operators also from the
viewpoint of their locality or nonlocality.
The mechanical quantities are local operators beyond any doubt.
On the other hand, whether the conserved quantities given 
by Eq.(\ref{def_px_cons})-Eq.(\ref{def_Lz_cons})
are local operators or not is a little delicate question owing to very special
nature of the Landau problem. This is because the quantity $B$
appearing in the definition Eq.(\ref{def_px_cons})-Eq.(\ref{def_Lz_cons}) 
can be considered a function either of the magnetic field at the point $\bm{x}$ or 
of the magnetic field at any other spatial point, because of the special feature 
that the strength of magnetic field is given by a constant $B$ everywhere. 
In that sense, we think that they should be distinguished from ordinary local 
operators like the mechanical momentum and OAM operators.
Probably, some additional explanation would be helpful to understand
this delicacy.

Let us consider the motion of electron under the magnetic
field, which is inhomogeneous but has axial symmetry around the $z$-axis
which goes through perpendicularly a predetermined origin in the $xy$-plane.
The magnetic field $\bm{B} (\bm{x})$ in this case is given by
$\bm{B} (\bm{x}) = B (r) \,\bm{e}_z$ with $r = \sqrt{x^2 + y^2}$. 
Accorging to \cite{Konstantinou2016},\cite{Konstantinou2017},
the conserved pseudo OAM operator (corresponding to our conserved 
OAM operator) is given by 
\begin{equation}
 \hat{L}^{cons}_z (\bm{A}) \ = \ \hat{L}^{mech}_z (\bm{A}) \ - \ e \, 
 \int^r_0 \,B (r^\prime) \,r^\prime \,d r^\prime.
 \label{inhomogeneous_OAM}
\end{equation}
It can be verified that it is in fact conserved and that it is a rotation
generator around the $z$-axis independently of the choice of $\bm{A}$. 
In a sense, it is a generalization of the
conserved OAM operator Eq.(\ref{inhomogeneous_OAM}) to the case
of inhomogeneous magnetic field.
The covariant transformation property of the conserved OAM operator defined
by Eq.(\ref{inhomogeneous_OAM}) is also obvious, since the mechanical OAM
operator in the 1st term transform gauge-covariantly, while the 2nd term 
expressed with the magnetic field is evidently gauge invariant. 
However, the 2nd term is
apparently nonlocal. It is also clear that, in the limit of homogeneous
magnetic field, i.e. $B (r^\prime) \rightarrow B$, the conserved OAM operator 
given by Eq.(\ref{inhomogeneous_OAM}) reduces to Eq.(\ref{def_Lz_cons}). 
This confirms the non-local nature of the 2nd term of Eq.(\ref{def_Lz_cons}),
although it would be a little hard to recognize it.

\vspace{1mm}
At any rate, different from the mechanical quantities, which are taken as 
manifestly local operators, the conserved momenta and the conserved OAM
operators in the Landau problem are not taken as genuinely local operators. 
Probably, this is not unrelated to the fact that 
the conserved quantities are not genuinely gauge-invariant quantity in
the sense that the expectation values of the conserved operators do not 
coincide between the Landau eigen-states belonging to idifferent 
gauge classes.
Exactly the same can be said for the g.c. extension of the canonical OAM of
quarks appearing in the nucleon spin decomposition problem.
The fact is that, despite the {\it formal} gauge-covariant transformation property, 
the g.c. extensions of the canonical OAM of quarks does not corresponds to
a gauge-invariant quantity in the most standard sense.
Under the presence of {\it nonzero} color-electromagnetic fields inside the 
nucleon, what is genuinely gauge-invariant is the mechanical OAM of quarks
not the canonical OAM or its g.c. extension. 

\section{Summary and Conclusion}
\label{sec7}

In the gauge-potential-independent formulation of the Landau problem, 
there exist three conserved quantities, i.e. two conserved momenta 
$\hat{p}^{cons}_x (\bm{A})$ and $\hat{p}^{cons}_y (\bm{A})$, and 
one conserved orbital angular momentum $\hat{L}^{cons}_z (\bm{A})$. 
The conservation of these quantities are guaranteed
not only in classical mechanics but also in quantum mechanics in the
sense that they commute with the Landau Hamiltonian $\hat{H} (\bm{A})$
for arbitrary choice of the gauge potential $\bm{A}$. 
However, these three operators
do not commute with each other. Then, if one wants to obtain the
eigen-functions of the Landau Hamiltonian, one 
must choose either of the following three options :  
\begin{itemize}
 \item[(1)]  construct simultaneous eigen-states of $\hat{p}^{cons}_x (\bm{A})$ 
 and $\hat{H} (\bm{A})$.
 \vspace{1mm}
 \item[(2)] construct simultaneous eigen-states of $\hat{p}^{cons}_y (\bm{A})$ 
 and $\hat{H} (\bm{A})$.
\vspace{1mm}
 \item[(3)]  construct simultaneous eigen-states of $\hat{L}^{cons}_z (\bm{A})$
 and $\hat{H} (\bm{A})$.
\end{itemize}
These three types of eigen-states are respectively characterized by the sets of 
quantum numbers $(n, k_x)$, $(n, k_y)$, and $(n, m)$, where $k_x$, $k_y$,
and $m$ are the eigen-values of the conserved operators $\hat{p}^{cons}_x (\bm{A})$,
$\hat{p}^{cons}_y (\bm{A})$, and $\hat{L}^{cons}_z (\bm{A})$ 
corresponding to the eigen-functions of the respective classes. 
In the traditional formulation of the Landau problem, these three eigen-states
are related to those of the 1st Landau gauge, the 2nd Landau gauge,
and the symmetric gauge, after suitably fixing the gauge potential configuration.   
In the gauge-potential-independent formulation, however, the construction of 
these three types of eigen-states can be done without fixing the form of 
the gauge potential at least formally.  This means that there are infinitely many 
eigenstates characterized by the quantum  numbers $(n, k_x)$, and similarly for
other two types of eigen-states.  

Still, a vitally important fact,
which we have emphasized in the present paper, is that the probability and 
current distributions of the Landau electron in the three types of
eigen-states are absolutely different. In fact, the probability and current 
distributions of the 1st type of eigen-states have a translational symmetry 
with respect to the $x$-axis,  those of the 2nd type eigen-states have a 
translational symmetry with 
respect to the $y$-axis, and those of the 3rd type eigen-states have a 
rotational symmetry with respect to the $z$-axis.
We emphasize that the statements above is true for {\it arbitrary choices}
of the {\it gauge potential configuration}, since the 
eigen-functions belonging to the same class are related by $U(1)$
gauge transformations or phase transformation. 
Importantly however, the eigen-functions belonging to different 
classes are not connected by a $U(1)$ gauge transformation or
a {\it phase transformation}. 
This is the very reason why the probability and current
distributions corresponding to the three different types of eigen-states
are totally different. For this reason, we claim in \cite{WH2022} that these 
three types of eigen-states fall into different gauge classes even in the 
gauge-potential-independent formulation of the Landau problem. 
This viewpoint was however criticized in \cite{Govaerts2023}, %Govaerts2023
and it was claimed that there is only one gauge class within
the gauge-potential-independent formulation of the Landau problem.

\vspace{1mm}
In our opinion, for confirming the validity of our viewpoint explained above, what 
plays a decisively important role is the comparison of the quantum expectation values 
of the above conserved quantities with those of the corresponding mechanical
quantities, i.e. the two mechanical momentum operators
$\hat{p}^{mech}_x (\bm{A})$ and $\hat{p}^{mech}_y (\bm{A})$, and the 
mechanical OAM operator $\hat{L}^{mech}_z (\bm{A})$ between the
different classes of eigen-states. 
Note that these mechanical quantities are standardly believed to be
manifestly gauge-invariant and observable quantities.
In fact, we have explicitly shown that the expectation values of
these mechanical operators between the above different gauge
classes of eigen-states coincide perfectly with each other,
which reconfirms that they are in fact {\it genuinely} gauge-invaiant quantities.
In sharp contrast, despite the novel covariant gauge transformation
properties of the conserved operators $\hat{p}^{cons}_x (\bm{A})$, 
$\hat{p}^{cons}_y  (\bm{A})$, and $\hat{L}^{cons}_z (\bm{A})$, we find that the
their expectation values in the three different classes
of eigen-states do not coincide with each other. This means that, in spite
of the novel covariant gauge transformation property, these conserved
quantities cannot be regarded as gauge-invariant quantities in the
standard sense. Hence, observations of the quantum numbers 
$k_x$, $k_y$, and $m$, which are the eigen-values of the
conserved operators $\hat{p}^{cons}_x (\bm{A})$, $\hat{p}^{cons}_y (\bm{A})$, and
$\hat{L}^{cons}_z (\bm{A})$ in the respective classes of eigen-states, 
would {\it contradict} the well-established gauge principle, even though they
are important quantum numbers characterizing the Landau eigen-states.
In fact, we have explicitly demonstrated nonobservable nature of the quantum 
number $k_x$ through the analysis of the familiar quantum Hall effect.
Similarly, nonobservable nature of the quantum number $m$ has been
demonstrated through the comparison of the Landau problem
and the problem of the 2-dimensional Harmonic oscillator.

\vspace{1mm}
To sum up, although the two conserved momentum operators and one 
conserved OAM operator in the Landau problem transform covariantly 
under arbitrary gauge tranformation just like the 
corresponding two mechanical momentum operators and 
the mechanical OAM operator do, there is a physically unmissable
difference between these quantities. 
We must say that, different from the mechanical quantities, 
much physical significance cannot be  given to the gauge-covariant 
transformation properties of the conserved quantities.  After all, they 
are not gauge-invariant quantities in the standard or physical sense.
Undoubtedly, exactly the same can be said for the 
gauge-covariant extension of the canonical OAM operator of quarks
appearing in the gauge-invariant decomposition problem of the 
nucleon spin. 

\vspace{1mm}
As a final summary, by making full use of the analytically solvable
nature of the Landau problem, we have clearly shown the existence
of two types of physical quantities in nature, which should be clearly 
descriminated from the perspective of gauge theory.
The first are quantities, which look seemingly gauge-invariant but actually not,
while the second are genuinely gauge-invariant quantities, which correspond 
to direct experimental observables. 
A critical difference between these two types of quantities follows from
the fact that the latter are related to the currents obtained from the minimal 
principle, while the former are not such quantities.

\vspace{4mm}
\noindent
{\bf Acknowledgments}

\vspace{2mm}
The present work has been achieved on the basis of a number of previous 
works, which were carried out in collaborations  
with Pengming Zhang, Yoshio Kitadono, Liping Zou, and also with
Akihisa Hayashi.  The author would like to thank all of them for
useful discussions. 

%Acknowledgments are not compulsory. Where included they should be brief. Grant or contribution numbers may be %acknowledged.

%Please refer to Journal-level guidance for any specific requirements.

%\section*{Declarations}

%Some journals require declarations to be submitted in a standardised format. Please check the Instructions for Authors %of the journal to which you are submitting to see if you need to complete this section. If yes, your manuscript must %contain the following sections under the heading `Declarations':

%\section*{Data Availability Statement :} %No Data associated in the manuscript

%\begin{itemize}
%\item There is no funding.
%\item Conflict of interest/Competing interests (check journal-specific guidelines for which heading to use)
%\item Ethics approval 
%\item Consent to participate
%\item Consent for publication
%\item No Data associated in the manuscript.
%\item Code availability 
%\item Authors' contributions
%\end{itemize}

%\noindent
%f any of the sections are not relevant to your manuscript, please include the heading and write `Not applicable' for that %section. 

\vspace{1mm}
%%%%%%%%%%%%%%%
%%%%%%%%%%%%%%%
%%%%%%%%%%%%%%%
%\begin{appendices}
\appendix

\section{Additional remarks on the relation between the canonical
and conserved quantities}

Some additional explanation on the relation between the canonical quantities
and the conserved (or pseudo) quantities would be helpful in order to
unravel nontrivial similarity and dissimilarity between them.
In a paper \cite{Berche2016} titled ``Gauge transformations and conserved 
quantities in classical and quantum mechanics'' , Berche et al. discuss three
possible scenarios where under gauge transformations : 
\begin{itemize}
\item[(i)] conservation laws are not preserved in the usual manner ;

\item[(ii)] non-gauge invariant quantities can be associated with
physical observables ; 

\item[(iii)] there are changes in the physical boundary conditions of the wave 
function that render it non-single-valued.
\end{itemize}

Our main concern here is the first two claims (i) and (ii). In particular, as noticed 
in the paper \cite{Yang2022} by Yang and MacDonald titled 
``Comments on Gauge Invariance'', if the above statement (ii) were correct, 
it would contradict the widely-believed gauge principle.
Here we reiterate the central claims by Berche as follows : 
{\it One must distinguish two kinds of physical quantities, both corresponds
to observables in quantum mechanics and represented by Hermitian operators.
The first ones are evidently gauge-invariant quantities in the sense that
they associate with the same quantity in different gauge (like position
and velocity) and can be simply measured and interpreted.
The second ones, like the canonical momentum or canonical angular
momentum, are not gauge-invariant.
They represent quantities that, being the generators of space-time
transformations, keep the same geometrical meaning but carry different
physical content in different gauges. Nevertheless, they might be
related to fundamental symmetries and they commute with the Hamiltonian
in the gauge where the Hamiltonian exhibits the total symmetry of
the system.} 

\vspace{2mm}
After all, they seem to state that, as long as one works in a particular
gauge, a gauge-dependent quantity corresponds to a ``physical observable''.
Undoubtedly, what is in their mind here are the familiar canonical
quantities like the canonical momentum and the canonical OAM.
As shown in our discussion on the Landau problem, we now know
the existence of far more sophisticated quantities than the canonical quantities.
They are the three conserved quantities, i.e. the conserved OAM 
$\hat{L}^{cons}_z$ and the two conserved momenta 
$\hat{p}^{cons}_x$ and $\hat{p}^{cons}_y$.
Different from the corresponding canonical quantities $\hat{L}^{can}_z$,
$\hat{p}^{can}_x$ and $\hat{p}^{can}_y$, the operators corresponding to the
conserved quantities commute with the Landau Hamiltonian 
with {\it arbitrary choice} of the gauge potential configuration, which means
that they are conserved irrespectively of the choice of gauge potential. 
Moreover, these three conserved operators transform covariantly
under arbitrary gauge transformation. Hence, one might have 
an impression that the existence of such quantities would give another 
support to the viewpoint advocated by Berche et al.
However, as argued in the present paper,
%as well as in some previous 
%literature \cite{WH2022, Govaerts2023}, 
there are three distinct classes of Landau eigen-states, which have
either of the following symmetry properties, i.e. the rotational symmetry
around the $z$-axis, the translational symmetry along the $x$-axis, and
the translational symmetry along the $y$-axis.
An assembly of eigen-states with the rotational symmetry is characterized
by the two quantum numbers $n$ and $m$, where $m$ is the eigen-value
of the conserved OAM operator $\hat{L}^{cons}_z$.
Here we recall the claim by Govaerts \cite{Govaerts2023} that 
this eigen-value $m$ corresponds to an observable, which is conjectured 
based on the covariant gauge-transformation property of the conserved OAM 
operator $\hat{L}^{cons}_z$.
Note however that this quantum number $m$ is also the eigen-value of the
{\it gauge-variant} canonical OAM operator $\hat{L}^{can}_z$ in the symmetric
gauge in the standard formulation. This logically means that, if 
the quantum number $m$ is in fact an observable, it amounts to saying that
the canonical OAM is also related to an observable. 
Exactly the same can be said for
the quantum numbers $k_x$ and $k_y$, which appear in the Landau eigen-states
having the translational symmetry along the $x$-axis and the $y$-axis.
At first glance, this consideration appears to match the claim by Berche et al. 
that the canonical quantities are in principle observable.
It is not correct, however. The truth is that both of canonical and conserved
quantities do not correspond to direct observables in conformity with
the widely-accepted common understanding. (Remember that the validity of this 
statement was already demonstrated through the analysis of the two concrete 
examples performed in sect.\ref{sec4} and sect.\ref{sec5}.), which explicitly shows
why the quantum numbers $m$ and $k_x$ characterizing the Landau eigen-states
do not correspond to direct experimental observables.)

\vspace{1mm}
To sum up, as was emphasized by Yang and MacDonald \cite{Yang2022},
what is meant by gauge invariance of some quantity
in quantum mechanics is that the quantum expetation values of the 
corresponding operator must coincide independently of the choice of gauge. 
Only quantum operators satisfying this property correspond to 
genuine or direct experimental observables.
As we have repeatedly emphasized in the present paper, there are 
three types of Landau eigen-states, which have
either of the following symmetry property, i.e. the rotation symmetry
around the $z$-axis, the translational symmetry along the $x$-axis, and
the translational symmetry along the $y$-axis. These tree types of
Landau eigen-states are assigned as belonging to three inequivalent
$U(1)$ gauge classes.
We have confirmed that the expectation values of the conserved OAM operator
$\hat{L}^{cons}_z$ between the different types of eigen-states do not coincide
with each other. The same is true also for the expectation values of
the conserved momentum operators $\hat{p}^{cons}_x$ and $\hat{p}^{cons}_y$.
This means that these three conserved quantities are not 
gauge-invariant quantities in the standard sense despite the 
gauge-covariant transformation property of the corresponding quantum 
operators, which in turn dictates that they do not correspond to direct 
observables contrary to the claim in \cite{Govaerts2023} and
\cite{Berche2016}. After all, the fact is that which theoretical formulation
of the Landau problem to choose, i.e. the traditional formulation or the
gauge-potential-independent formulation, does not change physical 
predictions verified through direct measurements.

\section{General formula relating the eigen-functions of the 2-dimensional
Harmonic oscillator \\ in the Cartesian basis and the spherical basis}\label{secB}

The relation between the eigen-functions of the 2-dimensional Harmonic
oscillator in the Cartesian-coordinate basis and the spherical-coordinate
basis is discussed in many textbooks of quantum mechanics. (See, for
example, Complement ${\tt D}_{\tt VI}$ in the book by Cohen-Tannoudji
et al. \cite{Cohen-Tannoudji2020}). 
However, it is not so easy to find an explicit general formula, which
relates the eigen-states in the two bases. The purpose of this
Appendix is to offer it.

The Hamiltonian of the 2-dimensional (isotropic) Harmonic oscillator is
given by
\begin{equation}
 H_{osc} \, = \, \frac{1}{2 \,m} \,( p^2_x \, + \, p^2_y ) \ + \ \frac{1}{2} \,m \, \omega^2 \,
 ( x^2 \, + \, y^2) .
\end{equation}
It is widely known that, if we introduce the ladder operators by
\begin{eqnarray}
 a_x &=& \frac{1}{\sqrt{2 \,m \,\omega}} \,( m \,\omega \,x \, + \, i \,p_x ), 
 \\ 
 a_y &=& \frac{1}{\sqrt{2 \,m \,\omega}} \,( m \,\omega \,y \, + \, i \,p_y ),
 \hspace{4mm}
\end{eqnarray}
they satisfy the following commutation relations (C. R.'s) :
\begin{eqnarray}
 [ a_x, a^\dagger_x ] = 1, \,  [ a_y, a^\dagger_y ]  =  1, \,  
 \mbox{other C.R.'s}  =  0, \hspace{5mm}
\end{eqnarray}
and the Hamiltonian can be expressed in the form
\begin{equation}
 H_{osc} \, = \, \omega \,( a^\dagger_x \,a_x + a^\dagger_y \,a_y + 1 ).
\end{equation}
The eigen-states of this Hamiltonian are also well-known. They are
specified by two non-negative integers $n_x$ and $n_y$ and satisfy
the following eigen-equation :
\begin{equation}
 H_{osc} \,\vert \,n_x, n_y \rangle \ = \ E \,\vert \,n_x, n_y \rangle ,
\end{equation}
with
\begin{equation}
 E \, = \, \left( n_x \,+ \, n_y \, + \, 1 \right) \,\omega .
\end{equation}
Here, the two quantum numbers $n_x$ and $n_y$ respectively corresponds 
to the numbers of oscillator quanta with respect to the $x$-direction and $y$-direction.
Th explicit form is the eigen-states $\vert n_x, n_y \rangle$ is given as
\begin{equation}
 \vert n_x, n_y \rangle \ = \ \frac{1}{\sqrt{n_x ! \,n_y !}} \,\,
 (a^\dagger_x)^{n_x} \,(a^\dagger_y)^{n_y} \,\,\vert 0, 0 \rangle ,
\end{equation}
where $\vert 0, 0 \rangle$ represents the vacuum of the oscillator
quanta in both directions. 
%We point out  that, by introducing the quantum
%number $n$ by $n = n_x + n_y$, the above eigen-energies can also be
%represented as $E = ( n + 1) \,\omega$. 

Another important quantity in the 2-dimensional Harmonic oscillator
is the (canonical) angular momentum operator defined by
\begin{equation}
 L_z \ \equiv \ x \,p_y \ - \ y \,p_x .
\end{equation}
With the use of the ladder operators, it can be expressed as
\begin{equation}
 L_z \ = \ i \,( a_x \,a^\dagger_y \ - \ a^\dagger_x \,a_y ) .
\end{equation}
As easily verified, it commutes with the (isotropic) Harmonic
oscillator Hamiltonian $H_{osc}$ : 
\begin{equation}
 [ L_z, H_{osc} ] \ = \ 0 .
\end{equation}
This means that there exist simultaneous eigen-states of $H$ and $L_z$.
To find out them, it is convenient to introduce the following ladder 
operators \cite{Cohen-Tannoudji2020} : 
\begin{eqnarray}
 a_+  = \frac{1}{\sqrt{2}} \,( a_x - i \,a_y ), \ \ \ 
 a_-  = \frac{1}{\sqrt{2}} \,( a_x + i \,a_y ) . \hspace{4mm}
\end{eqnarray}
They satisfy the following C.R.'s : 
\begin{eqnarray}
 \, [ a_+, a^\dagger_+ ] &=& \ 1, \  [ a_-, a^\dagger_- ] \ = \ 1 \\
 \,[ a_+, a_- ] &=& [ a_+, a^\dagger_- ] \, = \, [ a^\dagger_+, a_- ]
 \, = \, [ a^\dagger_+, a^\dagger_- ] \ = \ 0.
\end{eqnarray}
By using these ladder operators, $H_{osc}$ and $L_z$ are expressed as
\begin{eqnarray}
 H_{osc} &=& ( a^\dagger_+ \,a_+ \ + \ a^\dagger_- \,a_- \ + \ 1 ) \,\omega , \\
 L_z &=& a^\dagger_+ \,a_+ \ - \ a^\dagger_- \,a_- .
\end{eqnarray}
Now, the simultaneous eigen-states of $H_{osc}$ and $L_z$ are easily constructed as
\begin{eqnarray}
 \vert n_+, n_- \rangle = \frac{1}{\sqrt{n_+ ! \,n_- !}} \,
 (a^\dagger_+)^{n_+} \,(a^\dagger_-)^{n_-} \,\vert 0, 0 \rangle . \hspace{4mm} 
\end{eqnarray}
In fact, it can be shown that they satisfy the following eigen-equations : 
\begin{eqnarray}
 H_{osc} \,\vert n_+, n_- \rangle &=& E \,\vert n_+, n_- \rangle \nonumber\\
 \mbox{with} \ \ E &=& ( n_+ + n_- + 1) \,\omega,  \hspace{6mm} \\
 L_z \,\vert n_+, n_- \rangle &=& m \,\vert n_+, n_- \rangle \nonumber \\ 
 \mbox{with} \ \  m &=& n_+ - n_- . \hspace{5mm}
\end{eqnarray}
Introducing the quantum number $n$ by $n = n_+ + n_-$, the eigen-value
of $H_{osc}$ is given by $( n + 1) \,\omega$. On the other hand,
for a fixed integer $n$, the eigen-value of $L_z$
takes the following values : 
\begin{eqnarray}
 m &=& \ n_+ \ - \ n_- \ = \ 2 \,n \ - \ n_- \nonumber \\
 &=& 
 n, \,n - 2, \,\cdots , \,- (n - 2), \, - \,n,  \hspace{6mm}
\end{eqnarray}
which means that the state with energy $E = ( n + 1) \,\omega$ has
$(n + 1)$-fold degeneracy. Then, it is
convenient to characterize the simultaneous eigen-states
of $H_{osc}$ and $L_z$ by using the quantum numbers $n$ and $m$
instead of $n_+$ and $n_-$.
Such eigen-states satisfy the equations :  
\begin{eqnarray}
 H_{osc} \,\vert n, m \rangle &=& ( n + 1 ) \,\omega \,\vert n, m \rangle,  
 \hspace{4mm} \\ 
 L_z \,\vert n, m \rangle &=& \ m \,\vert n, m \rangle. 
\end{eqnarray}
Note that the states $\vert n, m \rangle$ correspond to the eigen-functions
solved in the 2-dimensional spherical coordinates, while
the states $\vert n_x, n_y \rangle$ correspond to
the eigen-functions solved in the 2-dimensional Cartesian coordinates.
Of our interest is a general formula, which relates these two types
of eigen-states.

\vspace{3mm}
\noindent
Using the relations $a_+ = \frac{1}{\sqrt{2}} \,( a_x - i \,a_y )$, and 
$a_- = \frac{1}{\sqrt{2}} \,( a_x + i \,a_y )$, we obtain
\vspace{-3mm}

%
%\begin{equation}
% \pmatrix{n \cr k}
%\end{equation}

\begin{eqnarray}
 \vert n, m \rangle &=& \frac{1}{\sqrt{\left( \frac{n+m}{2} \right) ! \,
 \left( \frac{n- m}{2} \right) !}} \,\,
 ( a^\dagger_+)^{\frac{n+m}{2}} \,\,\,
 ( a^\dagger_- )^{\frac{n-m}{2}} \,\,\vert 0, 0 \rangle \nonumber \\
 &=& \frac{1}{\sqrt{\left( \frac{n+m}{2} \right) ! \,
 \left( \frac{n- m}{2} \right) !}} \,\
 \frac{1}{\sqrt{2^{\frac{n+m}{2}} \,\,2^{\frac{n-m}{2}}}} \nonumber \\
 &\times& \sum_{k=0}^{\frac{n+m}{2}} \,
  \binom{\frac{n+m}{2}}{k} \,\,
 ( a^\dagger_x )^{\frac{n+m}{2} - k} 
 \,\, ( i \,a^\dagger_y )^k \nonumber \\
 &\times& \sum_{j=0}^n \,\binom{\frac{n-m}{2}}{j} \,\,
 ( a^\dagger_x )^{\frac{n+m}{2} - j} \,\,
 ( - \,i \,a^\dagger_y )^j \,\,\vert 0 \rangle_x \,\vert 0 \rangle_y .
\end{eqnarray}
After rearranging the equation, we eventually get
\begin{eqnarray}
 \vert n, m \rangle &=& \frac{1}{\sqrt{\left( \frac{n+m}{2} \right) ! \,
 \left( \frac{n- m}{2} \right) !}} \,\,\frac{1}{\sqrt{2^n}} \,\,\,
 \sum_{k=0}^{\frac{n+m}{2}} \,\,\sum_{j=0}^{\frac{n-m}{2}} \,\,
 \binom{\frac{n+m}{2}}{k} \,\binom{\frac{n-m}{2}}{j} \,\,
 i^{k-j} \hspace{15mm}  \nonumber \\
 &\,& \hspace{15mm} \times \sqrt{( n - k - j) ! \,(k + j) !}  \,\,\,
 \vert n - k - j \rangle_x \,\vert k + j \rangle_y . 
\end{eqnarray}
This is the general formula, which provides us with the relation between 
the $\vert n_x, n_y \rangle$-basis and $\vert n, m \rangle$-basis
in the problem of 2-dimensional Harmonic oscillator.

%\end{appendices}

%%===========================================================================================%%
%% If you are submitting to one of the Nature Portfolio journals, using the eJP submission   %%
%% system, please include the references within the manuscript file itself. You may do this  %%
%% by copying the reference list from your .mathbbl file, paste it into the main manuscript .tex %%
%% file, and delete the associated \verb+\bibliography+ commands.                            %%
%%===========================================================================================%%

%\bibliography{sn-bibliography}% common bib file

%% if required, the content of .bbl file can be included here once bbl is generated
%%\input sn-article.bbl

%% Default %%
%%\input sn-sample-bib.tex%

\end{document}